\documentclass[a4paper,11pt]{article}
\pdfoutput=1 % if your are submitting a pdflatex (i.e. if you have
\usepackage{amsmath}
\usepackage{jcappub} % for details on the use of the package, please
\usepackage{xcolor}
\definecolor{thecolor}{rgb}{0.1,0.5,1}
\usepackage{float}
\usepackage[T1]{fontenc} % if needed
\usepackage{amsmath}
\usepackage{scalerel}
\usepackage{tensor}
\usepackage{color, colortbl}
\definecolor{light-gray}{gray}{0.95}
\definecolor{TK}{rgb}{0.8,0,0}
\definecolor{darkred}{rgb}{0.7,0,0}

\usepackage[export]{adjustbox}

\newcommand{\p}{\partial}
\newcommand{\be}{\begin{equation}}  \newcommand{\ee}{\end{equation}}
\newcommand{\bea}{\begin{eqnarray}} \newcommand{\eea}{\end{eqnarray}}

\title{\boldmath The isotropic attractor solution of axion-SU(2) inflation: \\ Universal isotropization in Bianchi type-I geometry}

\author[\mathbf{a},1]{Ira Wolfson,\note{Corresponding author.}}
\author[\mathbf{b}]{Azadeh Maleknejad,}
\author[\mathbf{c}]{Tomoaki Murata,}
\author[\mathbf{d,e}]{Eiichiro Komatsu,}
\author[\mathbf{c}]{and Tsutomu Kobayashi.}

\affiliation[\mathbf{a}]{International School for Advanced Studies (SISSA), Data Science Excellence Department, Via Bonomea 265, 34136 Trieste, Italy
}

\affiliation[\mathbf{b}]{Theoretical Physics Department, CERN, 1211 Geneva 23, Switzerland}

\affiliation[\mathbf{c}]{Department of Physics, Rikkyo University, Toshima, Tokyo 171-8501, Japan}

\affiliation[\mathbf{d}]{Max Planck Institute for Astrophysics,\\ Karl-Schwarzschild-Str. 1, 85748 Garching, Germany
}

\affiliation[\mathbf{e}]{Kavli Institute for the Physics and Mathematics of the Universe (Kavli IPMU, WPI),\\ UTIAS, The University of Tokyo, Chiba, 277-8583, Japan
}

\emailAdd{iwolfson@sissa.it}
\emailAdd{azadeh.maleknejad@cern.ch}
\emailAdd{tmurata@rikkyo.ac.jp}

\emailAdd{komatsu@mpa-garching.mpg.de}
\emailAdd{tsutomu@rikkyo.ac.jp}

\abstract{SU(2) gauge fields coupled to an axion field can acquire an isotropic background solution during inflation. 
We study homogeneous but anisotropic inflationary solutions in the presence of such (massless) gauge fields. A gauge field in the cosmological background may pose a threat to spatial isotropy.
We show, however, that such models \textit{generally} isotropize in Bianchi type-I geometry, and the isotropic solution is the attractor. Restricting the setup by adding an axial symmetry, we revisited the numerical analysis presented in \cite{Wolfson:2020fqz}. We find that the reported numerical breakdown in the previous analysis is an artifact of parametrization singularity. We use a new parametrization that is well-defined all over the phase space. We show that the system respects the cosmic no-hair conjecture and the anisotropies always dilute away within a few e-folds.}

\begin{document}
\null\hfill\begin{tabular}[t]{l@{}}
  { CERN-TH-2021-076}
\end{tabular}\\
\null\hfill\begin{tabular}[t]{l@{}}
  { RUP-21-8}
\end{tabular}

\maketitle
\flushbottom

\section{Introduction}

Our Universe is nearly homogeneous and isotropic on cosmological scales. 
It is natural to seek a dynamical explanation for that unexpected symmetry, i.e., isotropic and homogeneous Universe is an
attractor solution of the cosmic evolution.  Gibbons and Hawking argued that the late-time behavior of any accelerating Universe is an isotropic Universe, i.e., ``cosmic no-hair conjecture'' \cite{Gibbons:1977mu,Hawking:1981fz}. 
Wald’s ``cosmic no-hair theorem'' proved that Bianchi-type models (except Bianchi type-IX) with a positive cosmological constant and a standard matter field would approach de Sitter space
exponentially fast \cite{Wald:1983ky}. The current data are in agreement with the concept of cosmic Inflation \cite{Starobinsky:1980te,Guth:1980zm,Linde:1981mu,Sato:1980yn,Albrecht:1982wi}, which postulates an epoch of quasi de Sitter expansion in the early Universe. Inflation, however, does not satisfy the conditions for Wald's theorem completely because it is driven by a rolling scalar field rather than a cosmological constant. Therefore, the inflationary version of the cosmic no-hair theorem states that (in the presence of spinning fields) anisotropies may grow during inflation, though their
amplitude is suppressed by the slow-roll evolution \cite{Maleknejad:2012as}.

While the particle physics of inflation is still unknown, one well-motivated candidate for the inflaton field is an axion field. Axions are naturally coupled to gauge fields which are the building blocks of particle physics models. Non-Abelian gauge fields may contribute to the physics of inflation and acquire a vacuum expectation value (VEV) while respecting the spatial isotropy \cite{Maleknejad:2011jw, Maleknejad:2011sq}.  Inspired by the original models (gauge-flation \cite{Maleknejad:2011jw, Maleknejad:2011sq} and chromo-natural inflation \cite{Adshead:2012kp}), several different realizations of SU(2)-axion inflation models have been proposed and studied in the literature (see \cite{Maleknejad:2012fw} and section 2 of \cite{Maleknejad:2018nxz} for reviews, and references therein). Gauge fields in physics of inflation give rise to a rich phenomenology. In particular, they produce particles during inflation, such as charged Higgs via the Schwinger effect \cite{Lozanov:2018kpk} and charged fermions by both the Schwinger effect \cite{Domcke:2018gfr,Maleknejad:2019hdr,Mirzagholi:2019jeb} and chiral anomaly \cite{Maleknejad:2020yys, Maleknejad:2021nqi}. As all the Sakharov conditions \cite{Sakharov:1967dj} are satisfied during inflation \cite{Maleknejad:2014wsa, Maleknejad:2016dci,Caldwell:2017chz,Adshead:2017znw}, it provides a natural setting for generating the matter-anti matter asymmetry \cite{Maleknejad:2020yys, Maleknejad:2021nqi}.
Another consequence of the Schwinger effect \cite{Maleknejad:2018nxz} is the sourced primordial gravitational waves. As a cosmological smoking gun, it predicts a stochastic background of chiral \cite{Maleknejad:2012fw, Adshead:2013qp, Dimastrogiovanni:2012ew,Maleknejad:2016qjz,Obata:2016tmo} and non-Gaussian \cite{Agrawal:2017awz,Agrawal:2018mrg,Dimastrogiovanni:2018xnn} primordial gravitational waves, which leads to parity-odd cross-spectra for CMB experiments and circular polarization for laser interferometers \cite{Thorne:2017jft}. Detection of this background is an excellent target for all gravitational wave experiments (CMB, pulsar timing arrays, and laser interferometers) across at least 21 decades in frequencies \cite{Campeti:2020xwn}. 

The SU(2) gauge field and its spatial isotropy have a number of compelling phenomenological and observational consequences. But is this isotropic gauge field's VEV the attractor solution? Do the SU(2)-axion models respect the cosmic no-hair conjecture? Embedding the gauge-flation and chromo-natural inflation models in Bianchi type-I geometry, the above questions have been addressed in \cite{Maleknejad:2011jr} and \cite{Maleknejad:2013npa}, respectively, and the case of massive SU(2) gauge field has been studied in \cite{Adshead:2018emn}. All these studies were based on assuming i) an axial symmetry in Bianchi type-I geometry and ii) that the SU(2) VEV is diagonal in the same frame as the metric. Based on these restrictive assumptions, it was shown that the massless SU(2) gauge fields coupled to the axion field by a Chern-Simons interaction do respect the cosmic no-hair conjecture in Bianchi type-I geometry. Therefore, the initial homogeneous but anisotropic geometrical deviations from the Friedmann–Lemaitre–Robertson–Walker (FLRW) metric are washed out during the period of inflation, and the gauge field's isotropic VEV is the attractor solution \cite{Maleknejad:2011jr, Maleknejad:2013npa}. However, in the case of massive gauge fields, the anisotropic solution can be the attractor if at least two colors of the gauge field take unequal masses \cite{Adshead:2018emn}.

These previous stability analyses have shortcomings; 1)
the assumption of axial symmetry,
and  2) restricting the numerical analysis to the limit that $\dot{A}_{\mu} \lesssim \mathcal{O}(1)H_0 A_{\mu}$. Here $A_{\mu}$ is the gauge field, a dot denotes time derivative, and $H_0$ is the Hubble expansion rate during slow-roll inflation.
In this paper, we address these issues for the massless case. Considering the most general Bianchi type-I geometry and homogeneous but anisotropic SU(2) field configurations, we prove that this setup always satisfies the cosmic no-hair conjecture. The key of the proof is that the Chern-Simons interaction with the axion only sources the isotropic part of the gauge field. We also extend the previous numerical analysis to the regime in which the kinetic term of the gauge field is large, i.e., $\dot{A}_{\mu} \gg H_0 A_{\mu}$. Recently, the authors of \cite{Wolfson:2020fqz} studied this regime for the spectator SU(2)-axion inflation model introduced in \cite{Dimastrogiovanni:2016fuu}. It was shown that the numerical analysis breaks down in that region, i.e., no-go area. In this paper, we show that this numerical breakdown was an artifact of our choice of parametrization, which is not well-defined around the no-go area. Considering a parametrization that is well-defined throughout the phase space, we study the system again, and the no-go area disappears. Therefore, the isotropic solution is the attractor of the inflationary solutions.

The rest of this paper is organized as follows. In section~\ref{sec:CN}, we study the chromo-natural inflation models \cite{Adshead:2012kp} in Bianchi type-I geometry. In section~\ref{sec:spectator}, we embed the spectator SU(2)-axion inflation model \cite{Dimastrogiovanni:2016fuu} in the same geometry and study the evolution of anisotropies and the VEV. In section~\ref{sec:Geometry}, we discuss the geometry of the anisotropic gauge field configuration. In section~\ref{sec-cosmic-no-hair}, we prove that the SU(2)-axion models with light gauge fields generally satisfy the cosmic no-hair condition in Bianchi type-I geometry. We conclude in section~\ref{sec:conclusions}. 

Throughout this work, we denote the variables at the initial time by the subscript $0$. We also work in natural units where $c=\hbar=1$, the reduced Planck mass $M_{Pl}$ is set to $1$, and the metric signature is the mostly positive one $(-,+,+,+)$.

\section{Chromo-natural model in Bianchi type-I geometry}\label{sec:CN}
The chromo-natural (CN) model for inflation is given by the action \cite{Adshead:2012kp}:
\begin{align}
    \mathcal{A}=\int d^4x \sqrt{-g} \left[-\frac{\mathcal{R}}{2} -\frac{\left(\partial_{\mu}\chi\right)^2}{2} - \mu^4\left(1+\cos\tfrac{\chi}{f}\right) -\frac{F^a_{\mu\nu}F_a^{\mu\nu}}{4} - \frac{\lambda \chi}{4f}\Tilde{F}^{a}_{\mu\nu}F_{a}^{\mu\nu} \right],
\end{align}
where $\mathcal{R}$ is the Ricci scalar, $\mu$ is the axion energy scale, $\lambda$ is the Chern-Simons coupling constant, and $F^a_{\mu\nu}$ is the field strength tensor of the $SU(2)$ gauge field given by
\begin{align}
    F^{a}_{\mu\nu}=\partial_{\mu}A^a_{\nu}-\partial_{\nu}A^a_{\mu}+g_{ \scriptscriptstyle{A}}\epsilon^a_{bc}A^b_{\mu}A^c_{\nu},
\end{align}
where $g_{ \scriptscriptstyle{A}}$ is the gauge coupling constant and $\epsilon^{a}_{bc}$ is the structure constant of the $SU(2)$ algebra.
Specifically, $A_\mu=A_\mu^aT_a$, 
where $\left\{T_a\right\}$ are the generators of the $SU(2)$ algebra with $a=1,2,3$ such that:
\begin{gather}
    T_a T_b =\frac{1}{4}\delta_{ab} I_2 + \frac{1}{2}i\varepsilon^{abc}T_c,
\end{gather}
in which $I_2$ is the $2\times 2$ identity matrix and $\varepsilon^{abc}$ is the totally antisymmetric matrix. The dual of the field strength tensor is given by:
\begin{align}
\Tilde{F}^{\mu\nu}_{a} \equiv \frac{\varepsilon^{\sigma\rho\mu\nu}}{2\sqrt{-g}}F^a_{\sigma\rho},
\end{align}
where $\varepsilon^{\sigma\rho\mu\nu}$ is the totally anti-symmetric tensor with $\varepsilon^{0123}=1$.

We embed this system in a Bianchi type-I geometry with axial symmetry in $x$-direction such that
\begin{align}
    ds^2=-dt^2 +e^{2\alpha(t)}\left(e^{-4\sigma(t)}dx^2 +e^{2\sigma(t)}\left(dy^2+dz^2\right)\right).
    \end{align}
 The Hubble expansion rate $H(t)$ is then given as
\begin{align}
H(t) \equiv \dot{\alpha}(t).
\end{align}
Upon introducing this geometry, we have the spatial triads for the SU(2) gauge group given by
\begin{align}
    e^a_1(t)=e^{\alpha-2\sigma}\delta^a_1,\; \quad e^a_{2}(t)=e^{\alpha+\sigma}\delta^a_2,\; \quad e^{a}_{3}(t)=e^{\alpha+\sigma}\delta^a_3.
\end{align}
To perform calculations, we use the temporal gauge for $A^a_{\mu}$:
\begin{align}
    A_0=0,\; A^a_i=\psi_i(t) e^a_i(t),
\end{align}
where $i$ runs from 1 to 3.
The axial symmetry is then set by $\psi_2=\psi_3$. We discuss the most general anisotropic geometry within Bianchi type-I and gauge field configurations in section~\ref{sec-cosmic-no-hair}. Following \cite{Maleknejad:2013npa}, one can decompose $\psi_i$ in terms of the isotropic and anisotropic components, $\psi$ and $\beta$, respectively, as 
\begin{align}
\psi_1(t) \equiv \frac{\psi(t)}{\beta^2(t)}, \quad \psi_2(t) \equiv \beta(t)\psi(t).\label{eq:Transformation}
\end{align}
The isotropic limit in the geometry and gauge field configuration ($\beta=\pm 1$) is given by
\begin{align}
\dot{\sigma}=0,  \quad  \psi_1(t)=\psi_2(t).
\end{align}

The matter Lagrangian of the model in terms of $\psi_{1}(t)$ and $\psi_2(t)$ is
\begin{align}
    \mathcal{L}_m &=\frac{\dot{\chi}^2}{2} -\mu^4\left(1+\cos \tfrac{\chi}{f}\right) + \psi_2^2\left(\dot{\alpha}+\dot{\sigma}+\frac{\dot{\psi_2}}{\psi_2}\right)^2+\frac{\psi_1^2}{2}\left(\dot{\alpha}-2\dot{\sigma}+\frac{\dot{\psi_1}}{\psi_1}\right)^2 \\ \nonumber  &-\frac{g_{ \scriptscriptstyle{A}}^2\psi_2^2}{2}\left(2\psi_1^2 +\psi_2^2\right)-\frac{3g_{ \scriptscriptstyle{A}} \lambda\chi \psi_1\psi_2^2}{f}\left[\dot{\alpha}+\tfrac{1}{3}\left(\tfrac{\dot{\psi_1}}{\psi_1}+\tfrac{2\dot{\psi_2}}{\psi_2}\right)\right],
\end{align}
where the last part is the Chern-Simons term, which does not contribute to the energy density. 

\subsection{Equations of motion}
The equations of motion for this system are given by the two Friedmann equations and the equations for the different fields. As the coordinate for the geometry anisotropy $\sigma$ has no potential term, we can assign
\begin{align}
  \mathcal{D}(t) \equiv  \frac{\partial \mathcal{L}}{\partial \dot{\sigma}}=2De^{-3\alpha(t)},
\end{align}
where $D$ is some integration constant \cite{Wolfson:2020fqz}. As $\mathcal{D}(t)$ is exponentially diluted by inflation, we ignore this term by setting it to zero for simplicity. This simplifies the study of other initial conditions. We thus have
\begin{align}
    \dot{\sigma}=\frac{\dot{\alpha}\left(\psi_1^2-\psi_2^2\right) +\dot{\psi_1}\psi_1-\dot{\psi_2}\psi_2}{3+2\psi_1^2 +\psi_2^2}. \label{Eq_sigma}
\end{align}

The energy density is divided into the axion term and the gauge field term:
\begin{align}
   \rho=\rho_{A}+\rho_{\chi},
\end{align}
with
\begin{align}
    \rho_{A}= \psi_2^2\left(\dot{\alpha}+\dot{\sigma}+\tfrac{\dot{\psi_2}}{\psi_2}\right)^2 + \frac{\psi_1^2}{2}\left(\dot{\alpha}-2\dot{\sigma}+\tfrac{\dot{\psi_1}}{\psi_1}\right)^2 +\frac{g_{ \scriptscriptstyle{A}}\psi_2^2}{2}\left(2\psi_1^2 + \psi_2^2\right),
    \end{align}
    and
    \begin{align}
    \rho_{\chi}=\frac{\dot{\chi}^2}{2}+\mu^4\left(1+\cos{\tfrac{\chi}{f}}\right).
    \end{align}
While it is possible to replace $\dot{\sigma}$ with the expression given in eq.~(\ref{Eq_sigma}), we leave $\rho_A$ in the aforementioned form for clarity.

The spatial part of the energy-momentum tensor is given as
\begin{gather}
T^i_{~j}(t) =  \delta^i_{~j} P(t) + \Pi^i_{~j}(t), 
\end{gather}
where $P(t)$ is the pressure and $\Pi^i_{~j}(t)$ is the anisotropic stress tensor, i.e., ${\rm Tr}[\Pi^i_{~j}(t)] = 0 $. 
The pressure $P$ is divided into the axion pressure $P_{\chi}$ and the gauge field pressure $P_A$ as
\begin{gather}
    P_{\chi}=\frac{\dot{\chi}^2}{2}-\mu^4\left(1+\cos{\tfrac{\chi}{f}}\right),\\
    \nonumber \\
    P_{A}=\frac{\rho_A}{3}.
\end{gather}
The anisotropic stress tensor can be written as  $\Pi^i_{~j}(t) = P_{A,aniso} ~ \rm{diag}(-2,1,1)$, where the anisotropic pressure is given by
\begin{gather}
    P_{aniso}=P_{A,aniso}= \frac{1}{3} \left[
  \psi_1^2 \left(\dot{\alpha}-2 \dot{\sigma} +\tfrac{\dot{\psi_1}}{\psi_1}\right)^2
  -\psi_2^2\left(\dot{\alpha}+\dot{\sigma}+\tfrac{\dot{\psi_2}}{\psi_2}\right)^2
  -g_{ \scriptscriptstyle{A}}\psi_2^2\left(\psi_1^2 -\psi_2^2\right)\right].
\end{gather}
Only the gauge fields contribute to the anisotropic pressure.

The equations of motion for the geometric anisotropy are given by the Friedmann equations:
\begin{gather}
    3\dot{\alpha}^2-3\dot{\sigma}^2=\rho,\label{Fried1}\\
    \nonumber \\
    \ddot{\sigma}+3\dot{\alpha}\dot{\sigma}=P_{aniso},\\
    \nonumber \\
    \ddot{\alpha}+3\dot{\sigma}^2=-\frac{\rho+P}{2}.
\end{gather}
The anisotropy in the geometry, i.e. $\dot{\sigma}$, is sourced by the anisotropic pressure.

The various fields' equations of motion are given by the principle of extremum action. This yields:
\begin{gather}
    \ddot{\psi}_1+3\dot{\alpha}\dot{\psi}_1=-\psi_1\left[2g_{ \scriptscriptstyle{A}}\psi_2^2 +2\left(\dot{\alpha}-2\dot{\sigma}\right)\left(\dot{\alpha}+\dot{\sigma}\right)+\left(\ddot{\alpha}-2\ddot{\sigma}\right)\right] +\frac{g_{ \scriptscriptstyle{A}}\lambda\dot{\chi}}{f}\psi_2^2,\\
    \nonumber \\
    \ddot{\psi}_2+3\dot{\alpha}\dot{\psi}_2=-\psi_2\left[ g_{ \scriptscriptstyle{A}}^2\left( \psi_1^2+\psi_2^2 \right)+\left( \dot{\alpha}+\dot{\sigma} \right) \left( 2\dot{\alpha}-\dot{\sigma} \right) \right]+\frac{g_{ \scriptscriptstyle{A}}\lambda\dot{\chi}}{f}\psi_1\psi_2,\\
    \nonumber \\
    \ddot{\chi}+3\dot{\alpha}\dot{\chi}=\frac{\mu^4}{f}\sin{\tfrac{\chi}{f}}-\frac{3g_{ \scriptscriptstyle{A}}\lambda\psi_1\psi_2^2}{f}\left[\dot{\alpha}+\tfrac{1}{3}\left(\tfrac{\dot{\psi_1}}{\psi_1}+\tfrac{2\dot{\psi_2}}{\psi_2}\right)\right].
\end{gather}
There is no direct correspondence between $\chi$ and $\dot{\sigma}$.

\subsection{Parameters and Initial conditions}\label{PIV-1}

In this work, we choose the scale of inflation to be $H_0=10^{-6}~M_{Pl}$, where the subscript $0$ always denotes an initial condition for the variables such as $\psi_0$, $\chi_0$, $\beta_0$, and $\dot\beta_0$. Assuming slow-roll inflation, this yields $\mu$ as
\begin{align}
    3H_0^2\simeq \mu^4\left(1+\cos\tfrac{\chi_0}{f}\right).\end{align}
In the isotropic limit, the equation of motion for $\chi$ in the slow-roll regime dictates:
\begin{align}
    \frac{\mu^4}{f}\sin{\tfrac{\chi}{f}}\simeq \frac{3g_{ \scriptscriptstyle{A}}\lambda H \psi^3}{f}.
\end{align}
We use this relation to set the initial value of $\psi_0$ in the isotropic limit as
\begin{align} 
    \psi_0 = \psi_{iso}=\left(\frac{\mu^4\sin{\tfrac{\chi_0}{f}}}{3g_{ \scriptscriptstyle{A}}\lambda H_0}\right)^{1/3}.\label{eq:psi_0}
\end{align}
In the isotropic limit (i.e. $\beta_0=\dot{\beta}_0=0$), the system is completely determined upon setting the values of $H_0,g_{ \scriptscriptstyle{A}},\lambda,f$ and $\chi_0$.  The parameters we use for this study are given in table~\ref{tab:parameters_1}.
\begin{table}[t]   
\centering
\begin{tabular}{lr}
    \begin{tabular}{||c|c||}
    \hline
        Parameter&Value\\
        \hline
        $\dot{\alpha}_0=H_0 $&$ 10^{-6}$  \\
        \hline
         $g_{ \scriptscriptstyle{A}}$&$2\times 10^{-3}$\\
         \hline
         $\lambda$& $200$\\
         \hline 
         $f$ & $0.01$\\
         \hline 
         $\chi_0$& $0.01\pi \times f$\\
         \hline
    \end{tabular} 
    &
    \begin{tabular}{||c|c||}
    \hline
        Parameter&Value\\
        \hline
        $\mu$ & $1.1\times 10^{-3}$ \\
         \hline
         $\psi_{iso}$&$\sim 3.4\times 10^{-3}$\\
         \hline
    \end{tabular}
    \end{tabular}
    \caption{The parameters and initial conditions we set to study the phase space in the CN system (left) and the parameters derived from these (right). The initial time derivatives of $\chi$ and $\psi$ are set to 0.}
    \label{tab:parameters_1}
\end{table}

\begin{figure}
    \centering
    \includegraphics[width=0.85\textwidth]{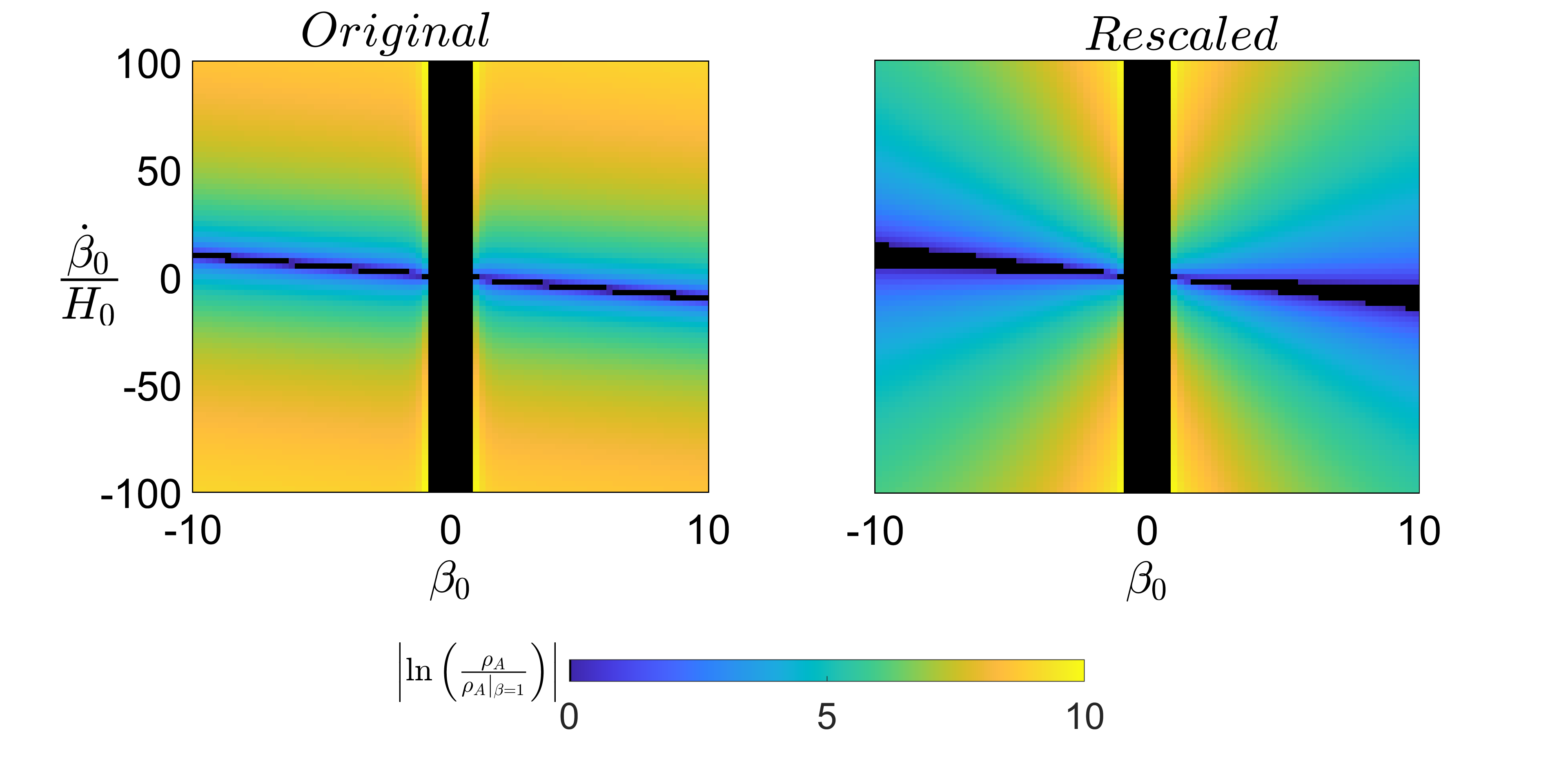}
    \caption{Applying the rescaling scheme in eq. \eqref{rescale-} enables a meaningful comparison between different initial condition sets. Without the rescaling (left panel), most of the phase space has an energy density several orders of magnitude higher than the energy density at $\beta=\pm 1$. After rescaling (right panel), most of the phase space  with $\dot\beta_0 \lesssim H_0 \beta_0$ becomes comparable to $\beta=\pm 1$ in terms of the energy density. The middle strip is masked, as the density there is very small ${\rho_A}/{\rho_{A, \beta=1}}\rightarrow 0$, hence $|\ln\rho_A/\rho_{A,\beta=1}| \rightarrow \infty$. }
    \label{fig:beta_scaling}
\end{figure}

\begin{figure}
    \centering
\includegraphics[width=0.5\textwidth]{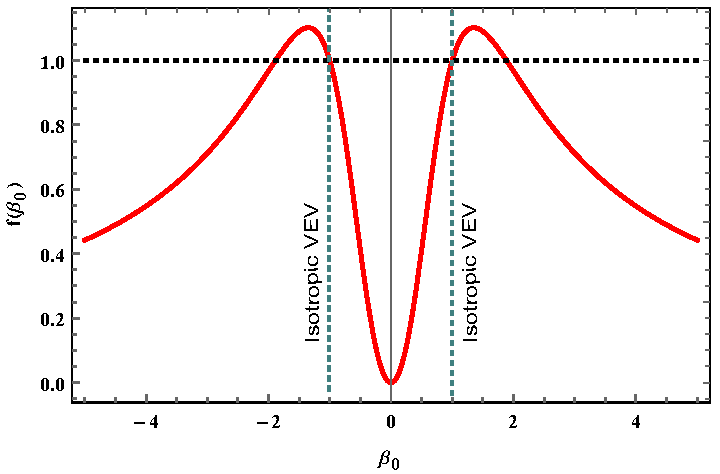}
    \caption{The rescaling function, $f(\beta_0)$ (eq.~\eqref{eq:fbeta0}), with respect to $\beta_0$. The points $\beta_0=\pm 1$ are the isotropic solutions for the VEV.}
    \label{fig:beta_scaling-0}
\end{figure}

In the anisotropic regime, the anisotropic parts of the gauge field contribute to the energy density and pressure. For a given $\psi_0$, the deviation of $\beta^2$ from one and that of $\dot\beta$ from zero increase the energy density of the system. In our phase space analysis, we are interested in studying anisotropic systems with (roughly) the same energy densities in the $\dot\beta_0 \lesssim   H_0 \beta_0$ region. That requires a $\beta$ dependent rescaling of $\psi_0$ comparing to the isotropic limit in eq. \eqref{eq:psi_0}, i.e.,
\begin{align}
 \psi_0 = f(\beta_0) \psi_{iso},
 \label{rescale-}
\end{align}
where $f(\beta_0)$ is worked out as follows.
The gauge field's energy density is given by
\begin{gather}
    \rho_{ \scriptscriptstyle{A}}=\frac{1}{2\beta^4}\left[\dot{\psi}+\psi\left(\dot{\alpha}-2\left(\dot{\sigma}+\tfrac{\dot{\beta}}{\beta}\right)\right)\right]^2 + \beta^2\left[\dot{\psi}+\psi\left(\dot{\alpha}+\left(\dot{\sigma}+\tfrac{\dot{\beta}}{\beta}\right)\right)\right]^2 +g_{ \scriptscriptstyle{A}}^2\frac{(\beta^6+2)\psi^4}{2\beta^2}.
\end{gather}
We consider the initial energy density of the system at different values of $\beta_0$ and $\dot{\beta}_0/H_0$. While there is a strong similarity in energy for $\beta\leftrightarrow -\beta$, the energy density away from the isotropic case ($\beta=\pm 1$) can be higher by orders of magnitude as seen in the left panel of figure~\ref{fig:beta_scaling}. Thus, when examining a set of initial conditions, we rescale the value of the initial {\it isotropic} part of the field $\psi_0$ to adjust the energy scale such that we have $\rho_A(\beta) \approx \rho_A\big|_{\beta=1}$ for $\dot\beta_0\lesssim H_0\beta_0$. This is accomplished by using the rescaling function:
\begin{gather}
    f(\beta_0)=\frac{\left(\tfrac{2n-1}{n}\right)\beta_0^2}{\left(1+\tfrac{n-1}{n}\left|\beta_0^3\right|\right)},
    \label{eq:fbeta0}
\end{gather}
%such that
%\begin{gather}
%    \psi_0 = f(\beta)\psi_{iso} = \frac{\left(\tfrac{2n-1}{n}\right)\beta_0^2}{\left(1+\tfrac{n-1}{n}\left|\beta_0^3\right|\right)}\psi_{iso},\label{rescale-}
%\end{gather}
with $n=5.3$ (see figure \ref{fig:beta_scaling-0}). With this rescaling, the point $\beta_0=0$ is associated with $\psi_0=0$ and vanishing gauge field's VEV,
\begin{equation}\label{no-VEV}
 \langle A_{\mu}^a (t)\rangle\rvert_{\beta_0=0} =0.   
\end{equation}
Therefore, the $\beta_0=0$  denotes the isotropic geometry with no gauge field in the background. This differs from the study in  \cite{Wolfson:2020fqz} without the rescaling, where $\beta_0= 0$ signifies a singularity point. With this rescaling, the initial energy density landscape becomes sufficiently regular, as seen in the right panel of figure~\ref{fig:beta_scaling}, which enables meaningful comparison of different initial conditions. The reason for disregarding the contribution of $\dot\beta_0$ in the rescaling of $\psi_0$ is that, for $\dot\beta_0 \gg H_0 \beta_0$, the axion interaction cannot support this large kinetic term of the gauge field, and it gets diluted away as radiation. 

\subsection{Phase space of anisotropic VEV}

In this section, we work out the phase space of the anisotropic part of the gauge field, i.e., ($\beta, \dot\beta$). Points in this phase space are associated with systems with the isotropic parameters and initial conditions given in table \ref{tab:parameters_1} while each given ($\beta,\dot\beta$) point represents the system with the same initial values of the anisotropic field. The value of $\psi_0$ is set as in \eqref{rescale-}. Given that, we study i) the number of e-folds it takes for each system to isotropize, and ii) whether the gauge field VEV isotropizes or dilutes away.    We use a numerical code that evaluates the trajectories of each system as a set of ordinary differential equations. We solve them using the Runge-Kutta (RK) method. We evaluate the next integration step with the 4th order RK method and assess the integration error with a 5th order RK method. The integration step is adjusted accordingly.
  
Since the equations of motion for $\psi_1,\psi_2$ and $\dot{\alpha}$ are intricately coupled, we use the following integration scheme. We construct the vector ${\bf V}=\left(\alpha,\psi_1,\psi_2\right)$ and for each coordinate we write the equation of motion in a bilinear form. We then construct
  and integrate a matrix equation of the form:
    \begin{gather}
        \ddot{V}_{i}=U_{j}M^{-1}_{ji} ,
    \end{gather}
    such that $\hat{M}=\hat{M}\left(q_i,\dot{q}_i\right)$ is a $3\times 3$ matrix of coordinates and their derivatives and $\bf U$ is a vector of potential type terms. This process is explained in detail in Appendix~\ref{Appendix:A}.     Most previous studies of this kind of a system employed e-folds as a de-facto clock. However, we use the cosmic time to avoid missing any features and constraining error propagation.

\begin{figure}[t]
    \centering   
    \includegraphics[width=0.9\textwidth,left=0.9\textwidth]{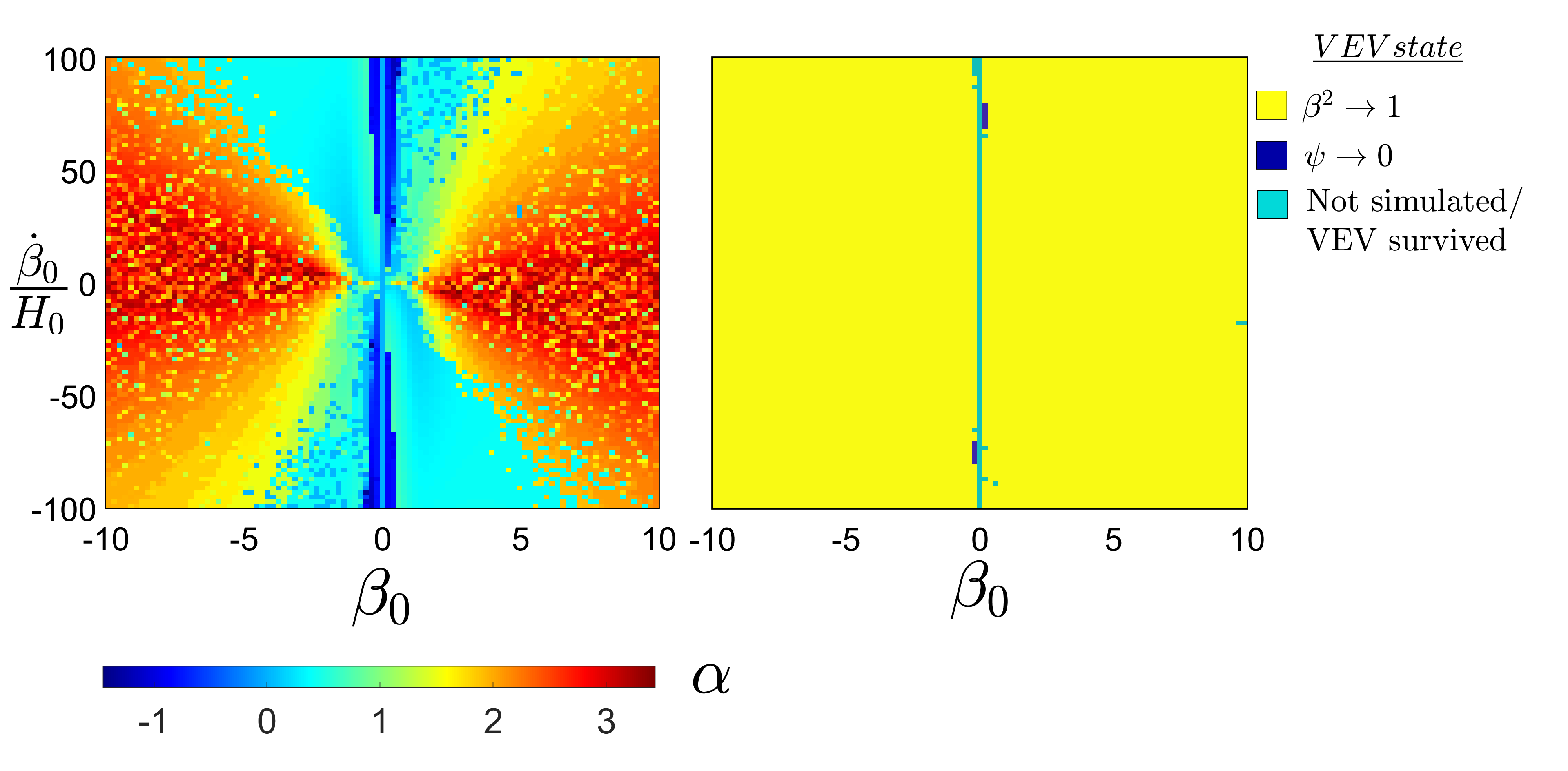} 
    \caption{The phase space of $(\beta_0,\dot{\beta}_0/H_0)$. Left panel: the absolute value of $\alpha$ in the color bar denotes the number of e-folds that takes for ${\dot\sigma}/{\dot\alpha}$ to become sufficiently small (see eq. \eqref{condition-1}). The negative $\alpha$ values show points for which the sign of $\beta(t)$ switches during its evolution. Without loss of generality and only for simplicity, here we assumed $\alpha_0=0$. Right panel: the phase space with focus on the evolution of the gauge field VEV. The colors mark points with final i) isotropization of the VEV (yellow), ii) dilution of the VEV during inflation (dark blue), and iii) points that are not simulated or the run time exceeded the expected limit (cyan). In both panels, the middle strip is not simulated. When $\beta=0$ while $\psi\neq 0$ the system is ill-defined. When $\psi=0$, it is FLRW by default. Thus the $\beta_0=0$ line is assigned the default value of $\alpha_0=0$.} \label{phase-space-CN}
\end{figure}

    With this numerical setup we evaluate each pair of the $(\beta_0,\dot{\beta_0})$ initial conditions, and first work out the number of e-folds it takes for the system to isotropize. We quantify the exit condition for the system as
    \begin{gather}\label{condition-1}
        \left|\frac{\dot{\sigma}}{\dot{\sigma}_0}\right|<10^{-3}\hspace{10pt}  \quad \textmd{and} \quad \hspace{10pt}\left|\frac{\dot{\sigma}}{\dot{\alpha}}\right|<10^{-3}.
    \end{gather}
 This condition will trigger the end of the simulation, provided that it is fulfilled for 100 consecutive simulation steps. This is done to avoid an erroneous exit due to some momentary or accidental instance of the conditions being met. The result of this simulation is presented in the left panel of figure~\ref{phase-space-CN}. 
 
 Next, to study the evolution of the gauge field VEV, we simulate the same system, but with different exit conditions, i.e.,
whether $\beta^2\rightarrow 1$:   \begin{gather}
(I) \quad    \hspace{10pt}\left|\beta^2-1\right|<10^{-3} \quad \textmd{and} \quad \frac{\left|\dot{\beta}\right|}{H_0}\ll 1 \quad  ( \textmd{where} \quad \frac{\left|\beta^2-1\right|}{\left|\beta_0^2-1\right|}<10^{-3}\hspace{10pt} ),\label{condition-2-1}
\end{gather}
 or the VEV dilutes away: 
 \begin{gather}
  (II) \quad  \left|\frac{\psi}{\psi_0}\right|<10^{-3} \quad \textmd{and} \quad \dot\psi<0.\label{condition-2-2}
\end{gather}
The result of this simulation is given in the right panel of figure~\ref{phase-space-CN}.

\begin{figure}[t]
    \centering
    \includegraphics[width=0.49\textwidth]{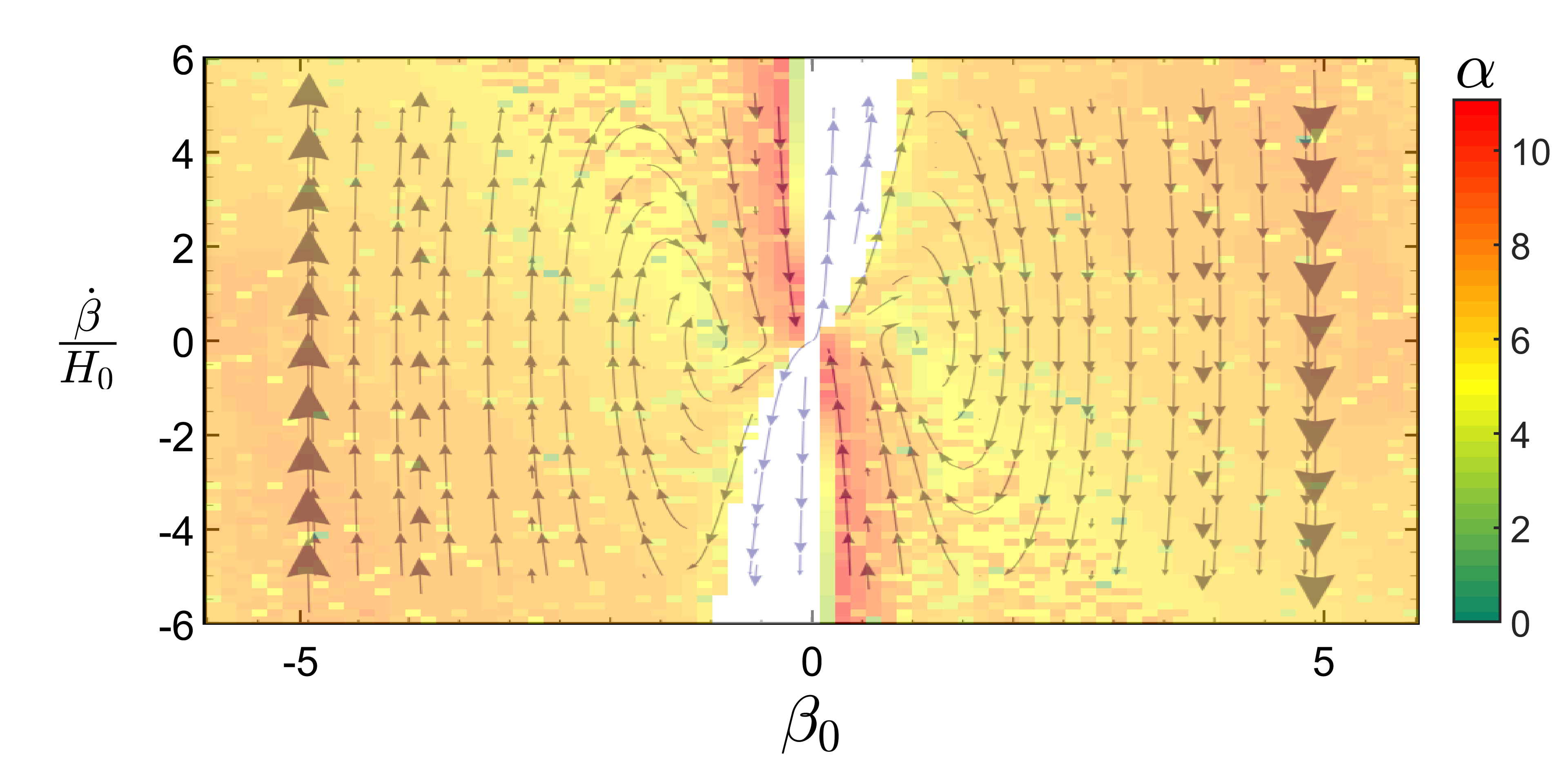}\includegraphics[width=0.49\textwidth]{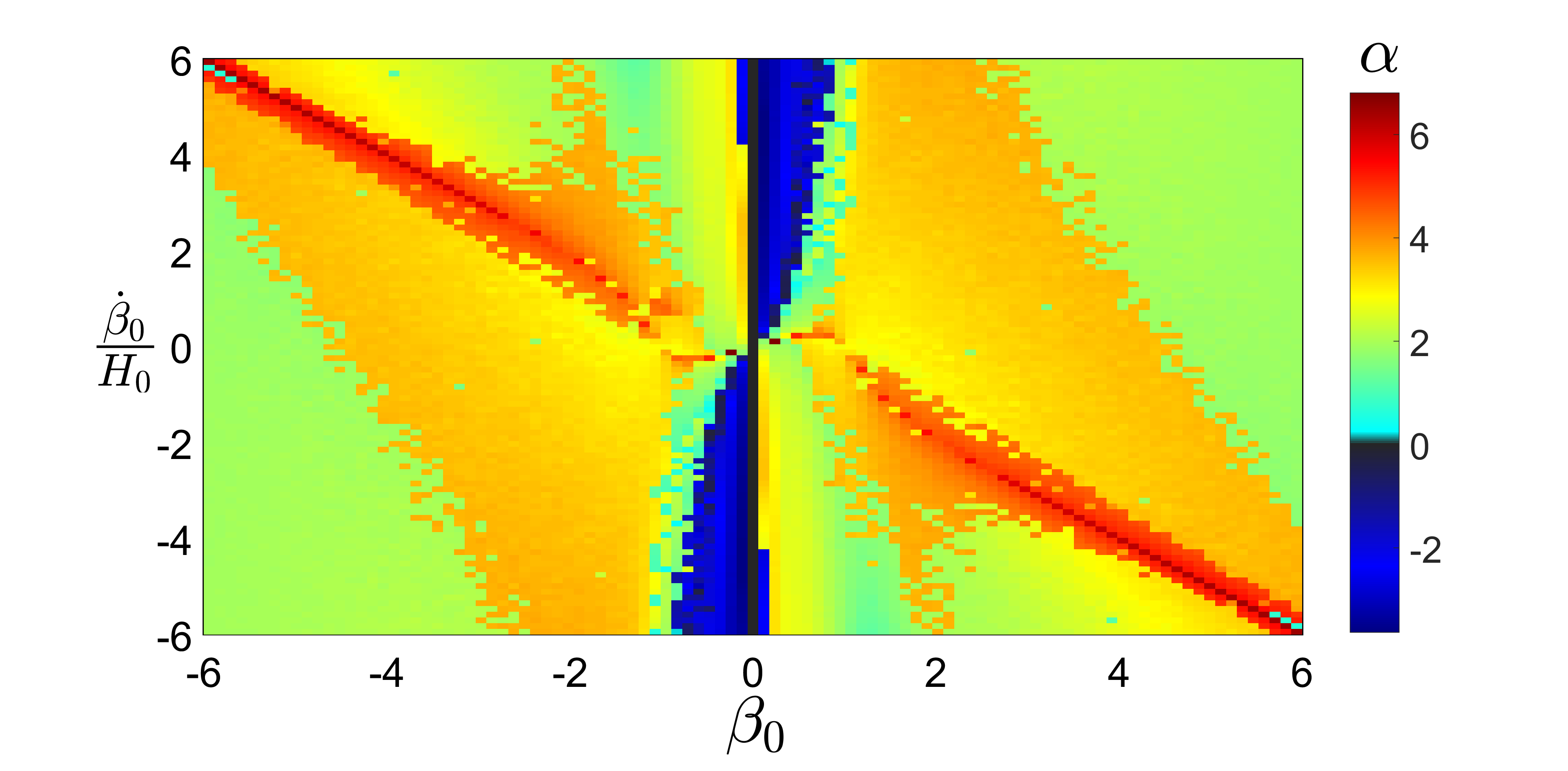}
    \caption{A comparison of the convergence picture in the ($\beta,\psi$) parametrization (left) and ($\psi_1,\psi_2$) parameterization (right).  The area previously failing to converge (the white region in the left panel) \cite{Wolfson:2020fqz} coincides with the region in which $\beta$ changes signs during the evolution. This plot shows the system studied in \cite{Wolfson:2020fqz} with the parameters given in table \ref{tab:parameters_22}.  The difference in e-folds-to-convergence is due to different exit conditions for $\dot{\sigma}/\dot{\alpha}$.}
    \label{fig:Comparison-0}
\end{figure}

   In \cite{Wolfson:2020fqz}, a similar setup but with different parameters (reproduced in table \ref{tab:parameters_22}; small gauge coupling, $g_{ \scriptscriptstyle{A}}=2\times 10^{-6}$ and $\lambda$ as large as 2000) has been studied in terms of $(\psi,\beta)$ parametrization for the VEV. It was shown that a sizable part of the parameter space was unstable, the so-called no-go area (see the left panel of figure~ \ref{fig:Comparison-0}). In our work, using a different parameterization for the VEV, i.e., $(\psi_1,\psi_2)$ instead of $(\psi,\beta)$, we study the same setup to understand the system in the no-go area. We find that: 1) the model isotropises in all of the parameter space, and 2) the points in the no-go area are associated with trajectories in which the $\beta$ field changes sign during its evolution (see the right panel of figure~\ref{fig:Comparison-0}). While the changing sign of $\beta$ is ill-defined in ($\psi,\beta$) parametrization, it is allowed in the ($\psi_1,\psi_2$) parametrization. Notice that when $\psi \neq 0$, the points $\beta=0$ and $\beta\rightarrow \pm \infty$ are the singularities of the system. Consequently, we see the entire phase space converges to the FLRW metric.

   \begin{table}[t]   
\centering
\begin{tabular}{lr}
    \begin{tabular}{||c|c||}
    \hline
        Parameter&Value\\
        \hline
        $\dot{\alpha}_0=H_0 $&$ 10^{-6}$  \\
        \hline
         $g_{ \scriptscriptstyle{A}}$&$2\times 10^{-6}$\\
         \hline
         $\lambda$& $2000$\\
         \hline 
         $f$ & $0.1$\\
         \hline 
         $\chi_0$& $0.01\pi \times f$\\
         \hline
    \end{tabular} 
    &
    \begin{tabular}{||c|c||}
    \hline
        Parameter&Value\\
        \hline
        $\mu$&$\sim 10^{-3}$  \\
         \hline
         $\psi_0$&$\sim \pi/200$\\
         \hline
    \end{tabular}
    \end{tabular}

    \caption{The parameters and initial conditions considered in \cite{Wolfson:2020fqz}.}    \label{tab:parameters_22}
\end{table}
   
\section{Spectator Axion-SU(2) system}\label{sec:spectator}
We now turn to the spectator version of the CN model introduced in \cite{Dimastrogiovanni:2016fuu}.
The action for this system is:
\begin{align}
    \mathcal{A}=&\int d^4x \sqrt{-g} \bigg[-\frac{\mathcal{R}}{2} -\frac{\left(\partial_{\mu}\chi\right)^2}{2} - \mu^4\left(1+\cos\tfrac{\chi}{f}\right) -\frac{F^a_{\mu\nu}F_a^{\mu\nu}}{4}\\
    \nonumber &\hspace{180pt}- \frac{\lambda \chi}{4f}\Tilde{F}^{a}_{\mu\nu}F_{a}^{\mu\nu} -\frac{\left(\partial_{\mu}\phi\right)^2}{2} -V(\phi)\bigg],
\end{align}
where $\phi$ is the inflaton field that dominates the energy density of the Universe at all times.
For later convenience, we define the ratio of potential energies as \cite{Wolfson:2020fqz}:
\begin{align}
    R=\frac{V(\phi)}{\mu^4\left(1+\cos\tfrac{\chi_0}{f}\right)},
    \end{align}
which is zero for the CN case while taking a large positive value for the spectator model. During the period of slow-roll inflation, $R\simeq {\rho_{\phi}}/{\rho_{CN}}$. 

\subsection{Parameters and initial values}
We set $H_0$, $\lambda$, $g_{ \scriptscriptstyle{A}}$, $f$ and $\chi_0$ similar to the CN model given in table \ref{tab:parameters_1}. Comparing with the CN model, the spectator axion-SU(2) inflation model has an extra parameter $R$. We find the values of $\mu$ and $\psi_0$ in this model as follows. Assuming slow-roll inflation, the Hubble expansion rate is given by
\begin{align}
    3H^2_0 \simeq\mu^4(1+\cos\tfrac{\chi_0}{f})+V(\phi_0),
    \end{align}
which, after using $\chi_0/f\ll 1$, yields $\mu^2\simeq \sqrt{\frac{3}{2}}\frac{H_0}{\sqrt{1+R}}$.

\begin{figure}[t]
    \centering
    \includegraphics[width=0.85\textwidth]{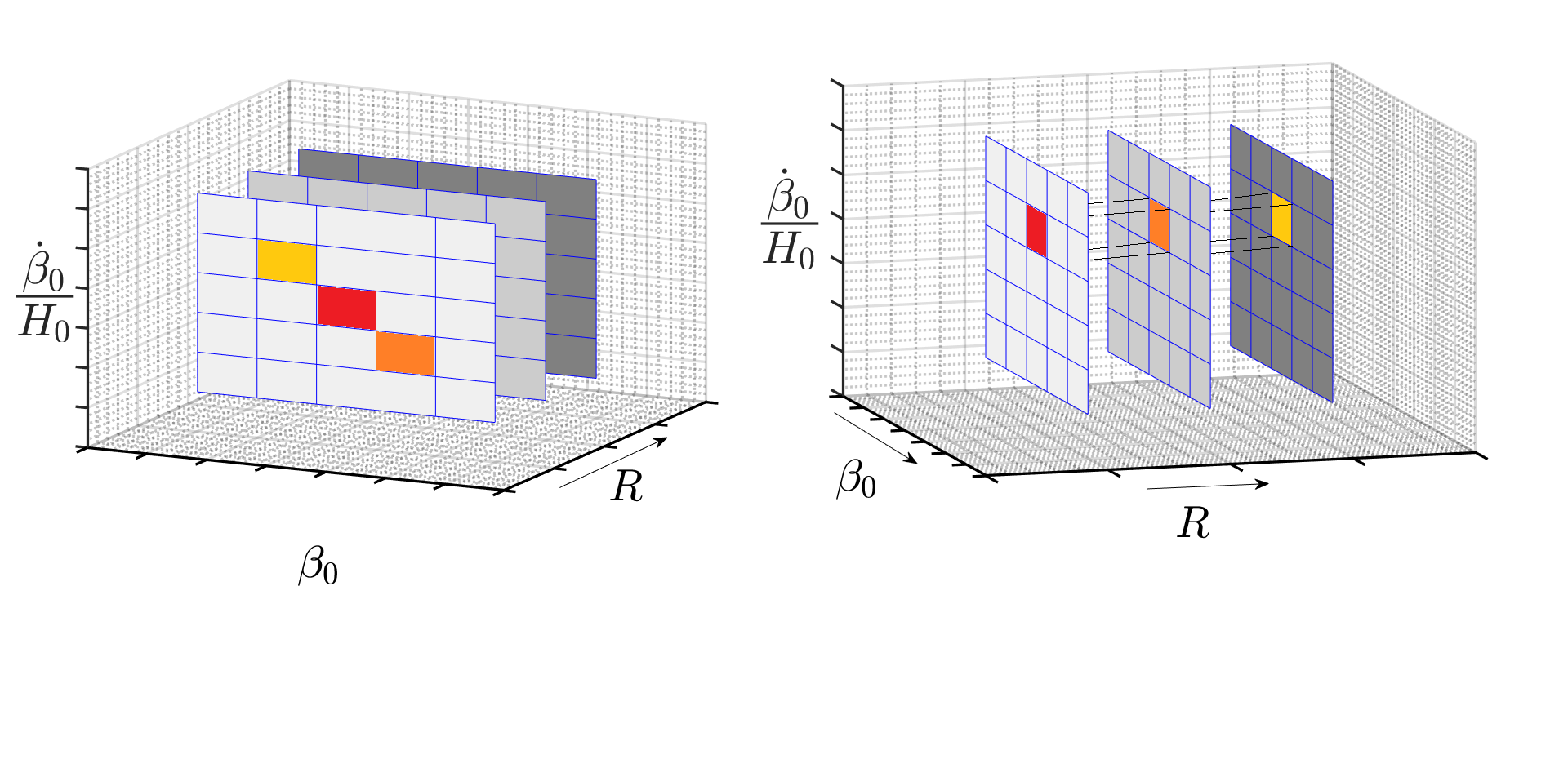}
    \caption{The different rescaling schemes of the phase space with respect to $\beta$ and $R$. In the left panel, the $\beta$-rescaling makes different cells in the same $R$-slice comparable, whereas in the right panel, the $R$-rescaling makes the same $\beta$-cell comparable for different $R$ values. }
    \label{fig:Scaling_B_R}
\end{figure}

\begin{figure}[t]
    \centering
    \includegraphics[width=0.5\textwidth]{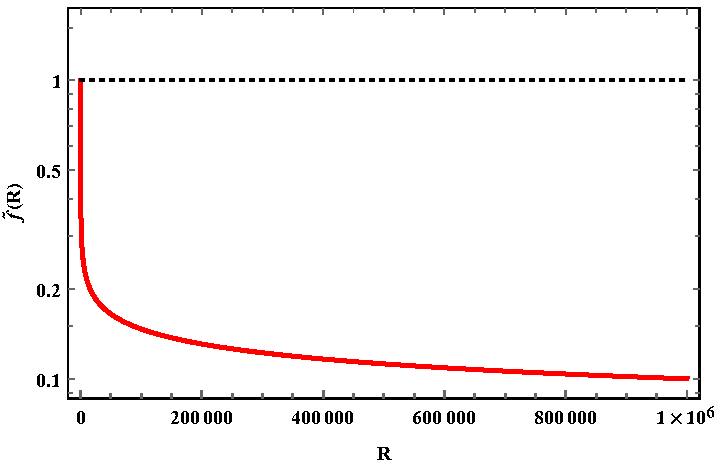}
    \caption{The rescaling function $\tilde{f}(R)$ with respect to $R$.}
    \label{fig:fR_rescaling}
\end{figure}

\begin{figure}[t]
    \centering
    \includegraphics[width=0.85\textwidth]{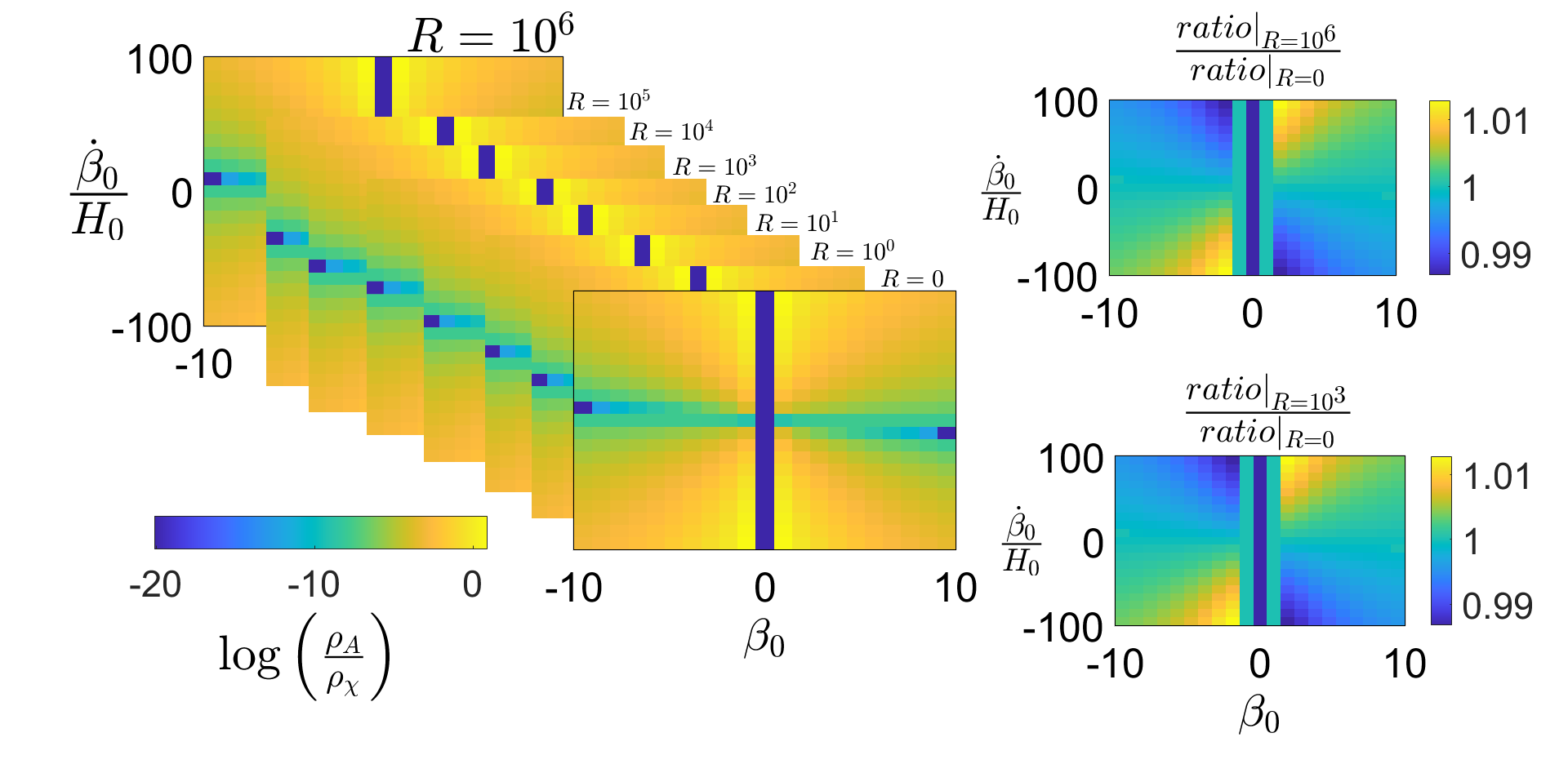}
    \caption{A set of initial $\rho_A$ to $\rho_{\chi}$ ratios across the studied phase space for different values of $R$ (left panels) and a detailed comparison between this ratio for $R=0$ and $R=10^3$ (lower right) and $R=10^6$ (upper right).}
    \label{fig:R_rescaling}
\end{figure}

As for the value of $\psi_0$, we want to study a comparable phase space for different values of $\beta_0$ and $R$. We worked out the proper rescaling for $\psi_{0}$ (comparing with the isotropic one) in section~\ref{PIV-1}. Here we find the proper rescaling between constant $R$ slices of phase space. This second rescaling is perpendicular to the rescaling for $\beta$, as illustrated in figure~\ref{fig:Scaling_B_R}.
This $R$-rescaling aims to keep the ratio ${\rho_{A}}/{\rho_{\chi}}$ constant across different $R$ values. Since in the isotropic limit for a given set of parameters (with fixed $H_0$) $\psi_0$ is completely given by $\mu$ (eq.~ \eqref{eq:psi_0}), we have $\psi_0\propto \mu^{4/3}$. The gauge field energy density is proportional to $\psi^2$, due to $g_A$ being small, and setting $\dot{\psi}_0\ll H_{0} \psi_0$.
Thus we have $\rho_A/\rho_{\chi}\propto \psi^2/\mu^4$.
Requiring a constant $\rho_A/\rho_{\chi}$ ratio, we recover $\psi\propto \mu^2$.
The rescaling function is given by
\begin{align}
    \psi_0\rightarrow \tilde{f}(\mu)\psi_0 \Rightarrow \tilde{f}(\mu)=\mu^{2/3}.
\end{align}
By virtue of $\mu^2\propto\tfrac{1}{\sqrt{1+R}}$ we can rewrite the above as $\mu^{2/3}\propto (1+R)^{-1/6}$,
arriving at a scaling law:
\begin{align}
    \psi_0\rightarrow \tilde{f}(R) \psi_0 = \left(1+R\right)^{-\tfrac{1}{6}} \psi_0.
\end{align}
This rescaling function is presented in figure~\ref{fig:fR_rescaling}.
Applying this rescaling function results in a fixed ${\rho_{A}}/{\rho_{\chi}}$ ratio to $1\%$ accuracy as evident in figure~\ref{fig:R_rescaling}.
Finally we apply the entire scaling scheme to both handle $\beta$ rescaling and $R$ rescaling:
\begin{align}
    \psi_0= \frac{\left(\tfrac{2n-1}{n}\beta_0^2\right)}{\left(1+\frac{n-1}{n}\left|\beta_0^3\right|\right)}\left(1+R\right)^{-\tfrac{1}{6}}\psi_{iso}. \label{rescale--}
\end{align}
\subsection{Phase space of anisotropic VEV} \label{sec:Numerical}

\begin{figure}[t]
    \centering
  \includegraphics[width=0.95\textwidth ,left=0.8\textwidth]{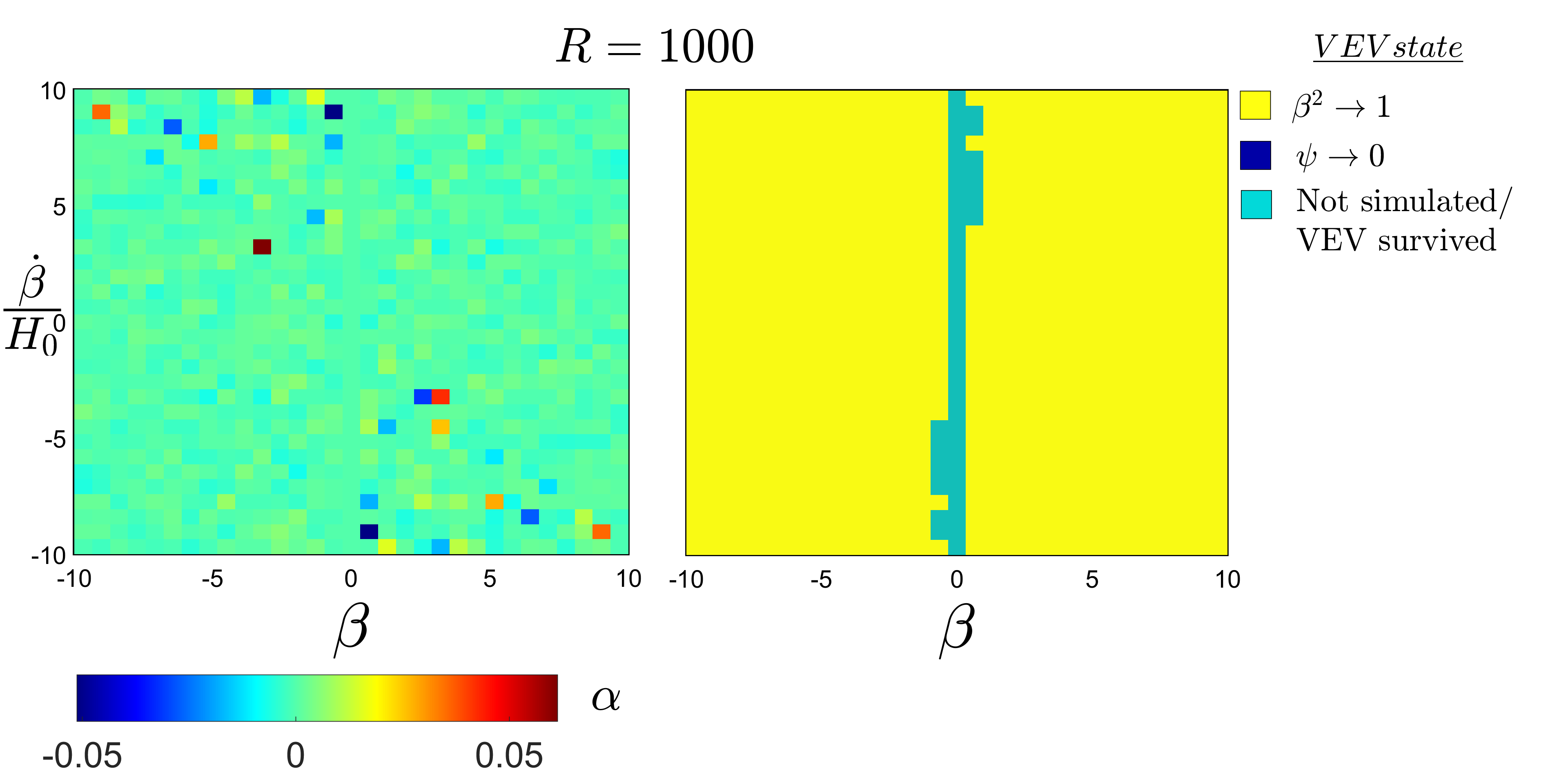}
    \caption{Same as figure~\ref{phase-space-CN} but for the spectator model with $R=1000$.}
    \label{fig:new_params_R}
\end{figure}

In this section we work out the phase space of the anisotropic part of the gauge field, i.e., ($\beta, \dot\beta$) for a given $R$. Points in this phase space are associated to systems with the isotropic parameters and initial conditions given in tables \ref{tab:parameters_1} and \ref{tab:parameters_22} while each given ($\beta,\dot\beta$) point represents the system with the same initial values of the anisotropic field. The value of $\psi_0$ is set as in \eqref{rescale--}. Similar to the previous section, we then study i) the number of e-folds it takes for each system with a given $R$ to isotropize (based on the conditions given in \eqref{condition-1}), and ii) whether the gauge field VEV isotropizes or dilutes away (based on the conditions given in \eqref{condition-2-1}-\eqref{condition-2-2}). One unfortunate side effect of setting $g_{ \scriptscriptstyle{A}}\sim 10^{-3}$ (parameter set in table \ref{tab:parameters_1}) comparing with $g_{ \scriptscriptstyle{A}}\sim 10^{-6}$ (parameter set in table \ref{tab:parameters_22}) is a longer computational runtime that scales with $R$ for each simulation. To account for that we reduced the resolution from $100\times 100$ pixels at $R=0$ to $30 \times 30$ at $R=10^3$. This accounts for the different resolutions between figures \ref{phase-space-CN} and \ref{fig:new_params_R}.

\begin{figure}[t]
    \centering
    \includegraphics[width=0.95\textwidth]{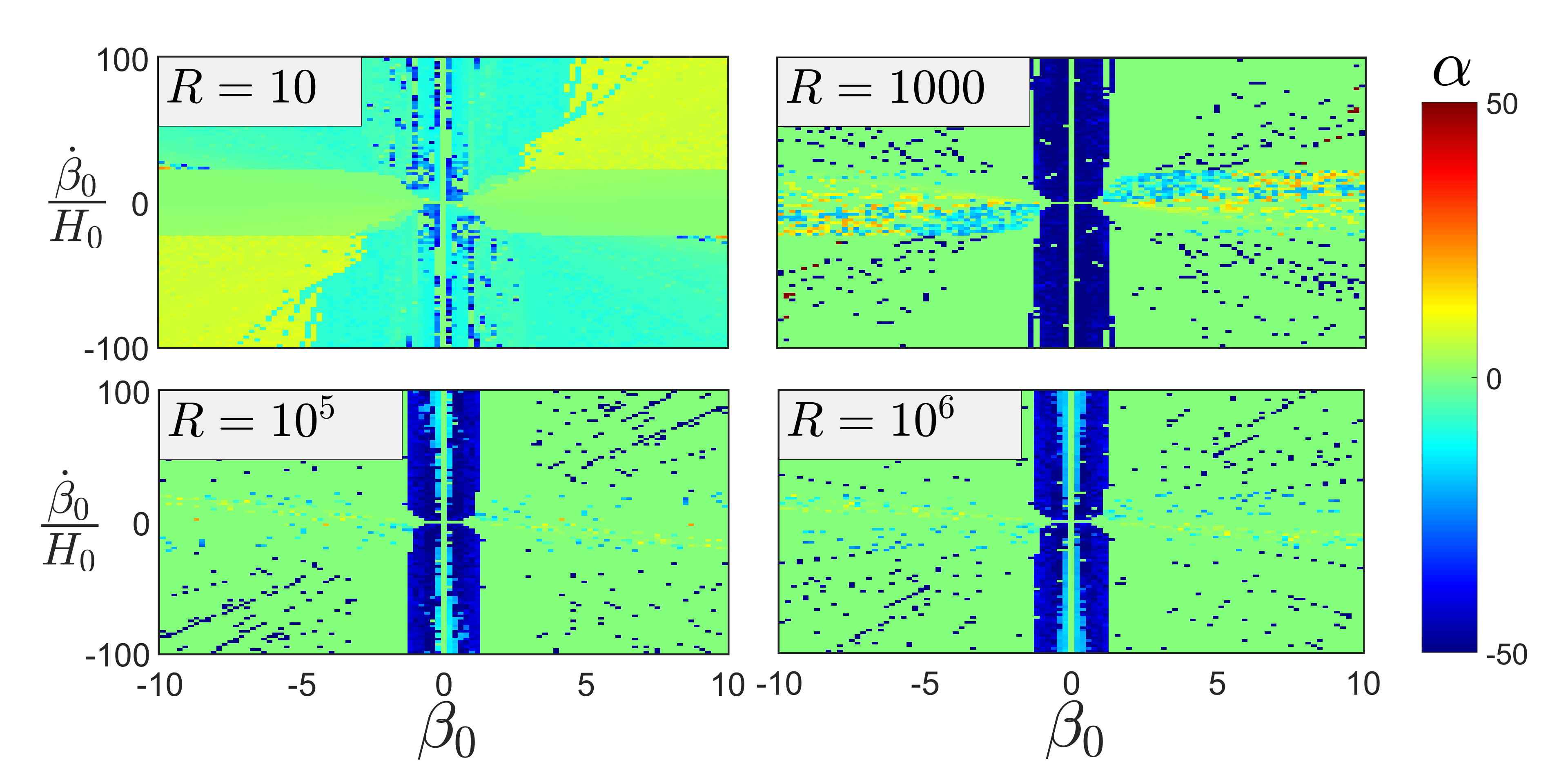}
    \caption{The number of e-folds required for the spectator system to converge to the FLRW metric (configuration given in table \ref{tab:parameters_22}). The entire phase space converges, with the number of required e-folds decreasing for greater dominance of the inflaton potential (larger $R$). However, the region between $\beta\in(-1,1)$ takes $\sim 50$ e-folds to converge, independent of $R$. }
    \label{fig:g_6_new_alpha}
\end{figure}

The result of these simulations for parameters given in table \ref{tab:parameters_1} and with $R=1000$ is shown in figure \ref{fig:new_params_R}. It shows that the VEV is stable, and the system isotropizes much faster than the CN setup. More precisely, for CN with $R=0$ (see figure~\ref{phase-space-CN}), the required e-folds is around 3.5, while for $R=1000$ this number decreases by $\sim 2$ orders of magnitude. For the parameters given in table \ref{tab:parameters_22} and for several values of $R$, the result of simulations for the number of e-folds required for isotropizations and the stability of the VEV  are presented in figure \ref{fig:g_6_new_alpha} and \ref{fig:psi_beta_map:g=10^{-6}}, respectively. Again the system always isotropizes regardless. With such a low gauge coupling constant, i.e., $g_{ \scriptscriptstyle{A}}\sim 10^{-6}$, however, the VEV is not stable but dilutes away in most of the phase space. The larger the $R$ is, the larger the region of the VEV's dilution becomes.

    \begin{figure}[t]
    \centering
    \includegraphics[width=0.95\textwidth]{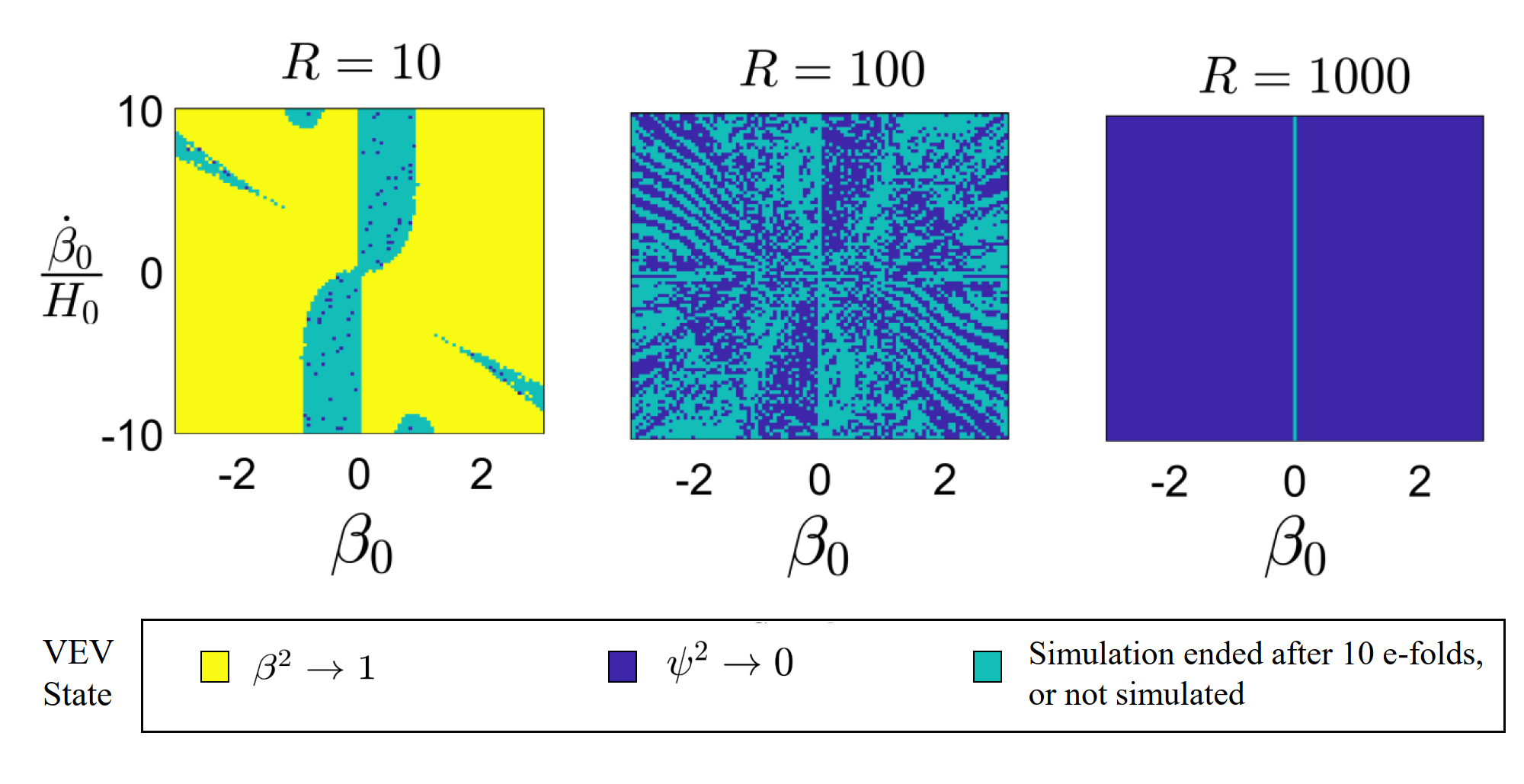}
    \caption{The final configuration of gauge fields  for systems with $g_{ \scriptscriptstyle{A}}\sim 10^{-6}$ (configuration given in table \ref{tab:parameters_22}), for different values of $R$. For the pure CN system ($R=0$), the gauge fields for most of the phase space converge on the non-trivial VEV, or survive the first 10 e-folds of inflation. Here, however, we observe a transition to a vanishing VEV as $R$ increases.} \label{fig:psi_beta_map:g=10^{-6}}
\end{figure}

\section{Geometry of the anisotropic gauge field configuration}\label{sec:Geometry}

In this section we compare the geometry of the SU(2) gauge field's VEV in terms of $(\psi,\beta)$ and $(\psi_1,\psi_2)$ parametrizations. This system was numerically studied recently using the $(\psi,\beta)$ parametrization  \cite{Wolfson:2020fqz}. It showed that in part of the anisotropic phase space with large values of ${\dot\beta_0}/({H_0\beta_0})$, the equations of motion failed due to a runaway effect of a kinetic term in the gauge fields energy density. That area in the phase space was previously called the no-go area. In sections \ref{sec:CN} and \ref{sec:spectator}, we demonstrated that the no-go area's apparent instability was a numerical issue which should be avoided by working in terms of the $(\psi_1,\psi_2)$ parametrization. More precisely, the spectator SU(2)-axion model (and hence CN) always isotropizes in a few e-folds. However, the VEV of the gauge field can have more complicated dynamics, which could only be captured in the $(\psi_1,\psi_2)$ parametrization. In this section, we further clarify the evolution of this anisotropic VEV in the previously called no-go area.

\begin{figure}[t]
    \centering
    \includegraphics[width=0.95\textwidth]{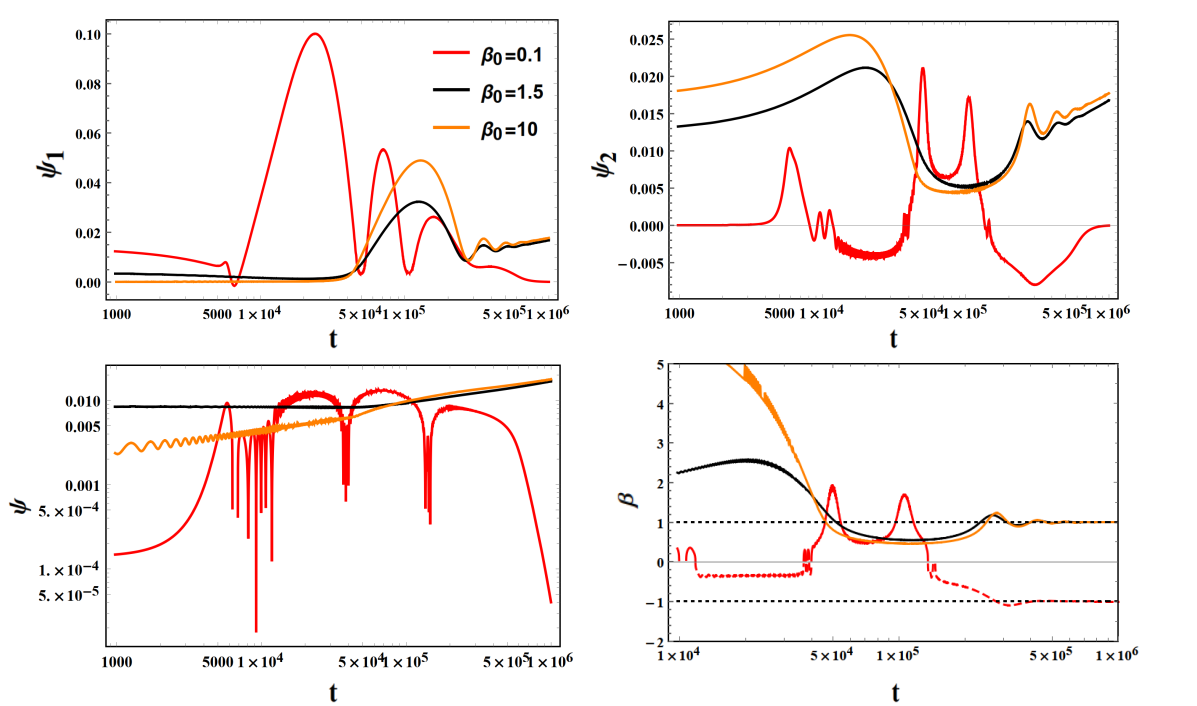}\caption{The CN model with the parameters given in table \ref{tab:parameters_1} and ${\dot\beta_0}/({H_0\beta_0})=10$, for three different values of $\beta_0$. The system is solved in terms of the $(\psi_1,\psi_2)$ parametrization and $t=2\times 10^{-5}$ corresponds to $N=1$ e-fold. The upper left and right panels show $\psi_{1}$ and $\psi_2$ vs time, while the lower left and right panels show the same system in terms of the $\psi$ and $\beta$ parametrization. For $\beta_0=0.1$, the $\beta$ field switches sign and it is continuous at $\beta=0$. } \label{fig:beta_psi_2}
\end{figure}

Figure \ref{fig:beta_psi_2} shows the $R=0$ system (CN model) with the parameters given in table \ref{tab:parameters_1} for three different values of $\beta_0$ and a large value of ${\dot\beta_0}/({H_0\beta_0})=10$. We find that the model  isotropizes in all of these cases within a few e-folds. In the $\beta_0=0.1$ case, however, $\beta$ passes zero and changes sign. In this system the isotropic VEV is not stable and it dilutes away by the expansion of the Universe. 

\begin{figure}[t]
    \centering
    \includegraphics[width=0.95\textwidth]{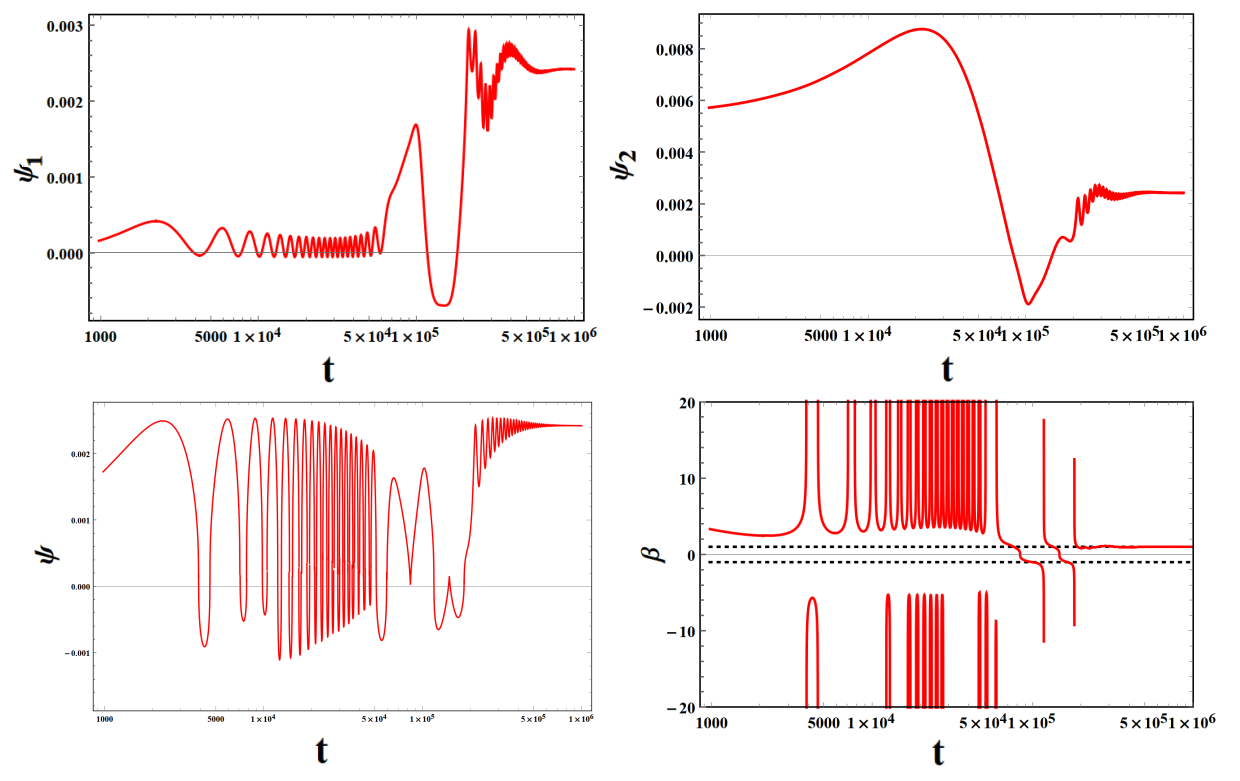}\caption{The spectator model with $R=10^{3}$, $\beta_0=10$, and ${\dot\beta_0}/({H_0\beta_0})=10$, for the parameters given in table \ref{tab:parameters_1}. The system is solved in terms of the $(\psi_1,\psi_2)$ parametrization. The upper left and right panels show $\psi_{1}$ and $\psi_2$ vs time, while the lower left and right panels show the same system in terms of the $\psi$ and $\beta$ parametrization. Here the $\beta$ field switches sign but it is \it{discontinuous} at $\beta=0$. } \label{fig:beta_psi}
\end{figure}

Figure \ref{fig:beta_psi} shows the spectator model with $R=10^{3}$, $\beta_0=10$, and ${\dot\beta_0}/({H_0\beta_0})=10$. We find that the $\beta(t)$ field switches sign but it is {\it{discontinuous}} at $\beta=0$. Working in terms of ($\psi,\beta$), the changing sign of $\beta(t)$ is analytically prohibited, and numerically unstable. Thus it gives rise to the notion of $\beta=0$ being a separatrix. Working in terms of $(\psi_1,\psi_2)$, on the other hand, the changing sign of $\beta$ is allowed. In the following, we will discuss the geometry of these two kinds of behavior of gauge field's VEV in passing through $\beta=0$ point.

\subsection{The geometry of parametrization}

\begin{figure}[t]
    \centering
    \includegraphics[width=0.45\textwidth]{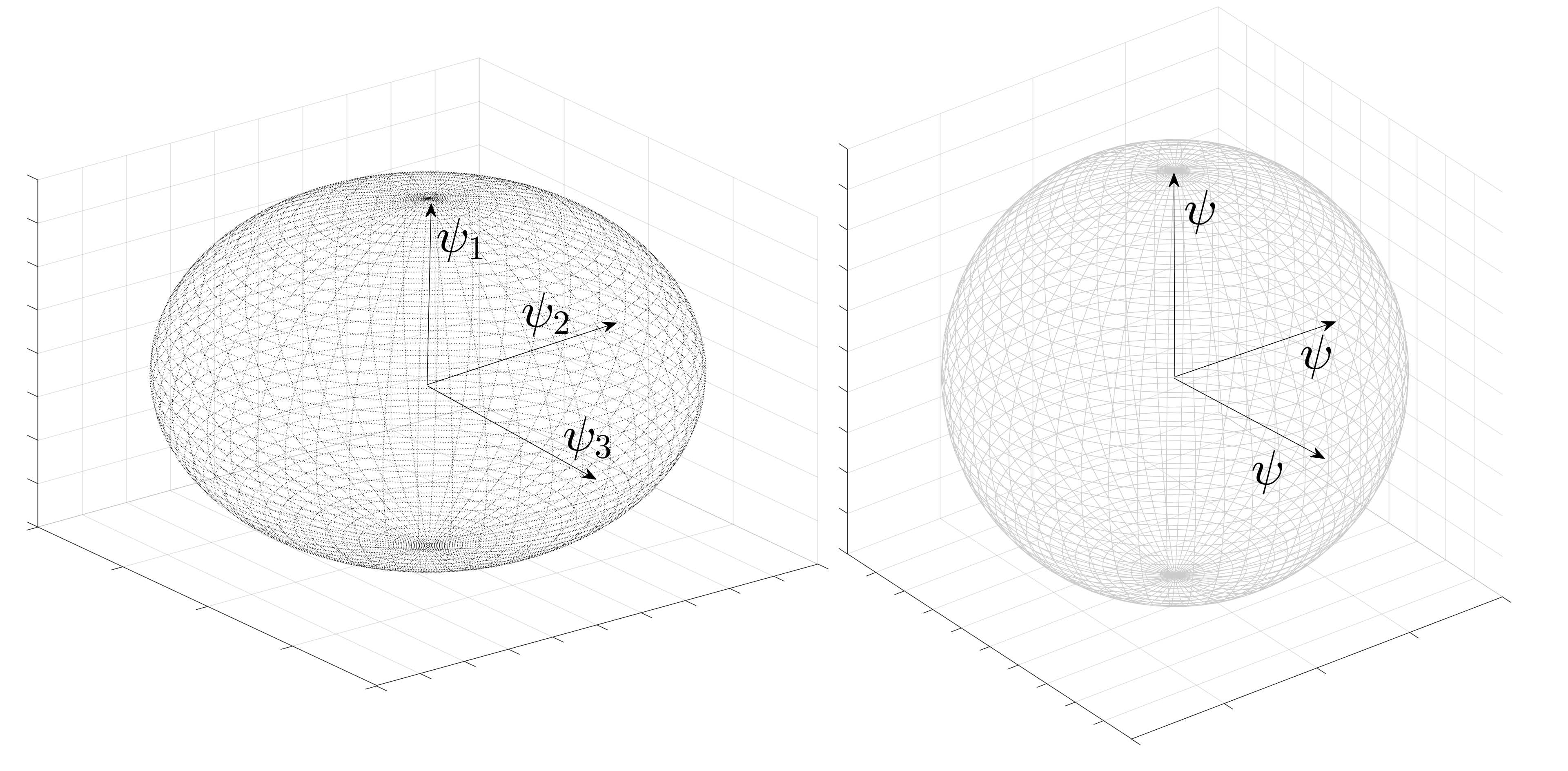}\includegraphics[width=0.45\textwidth]{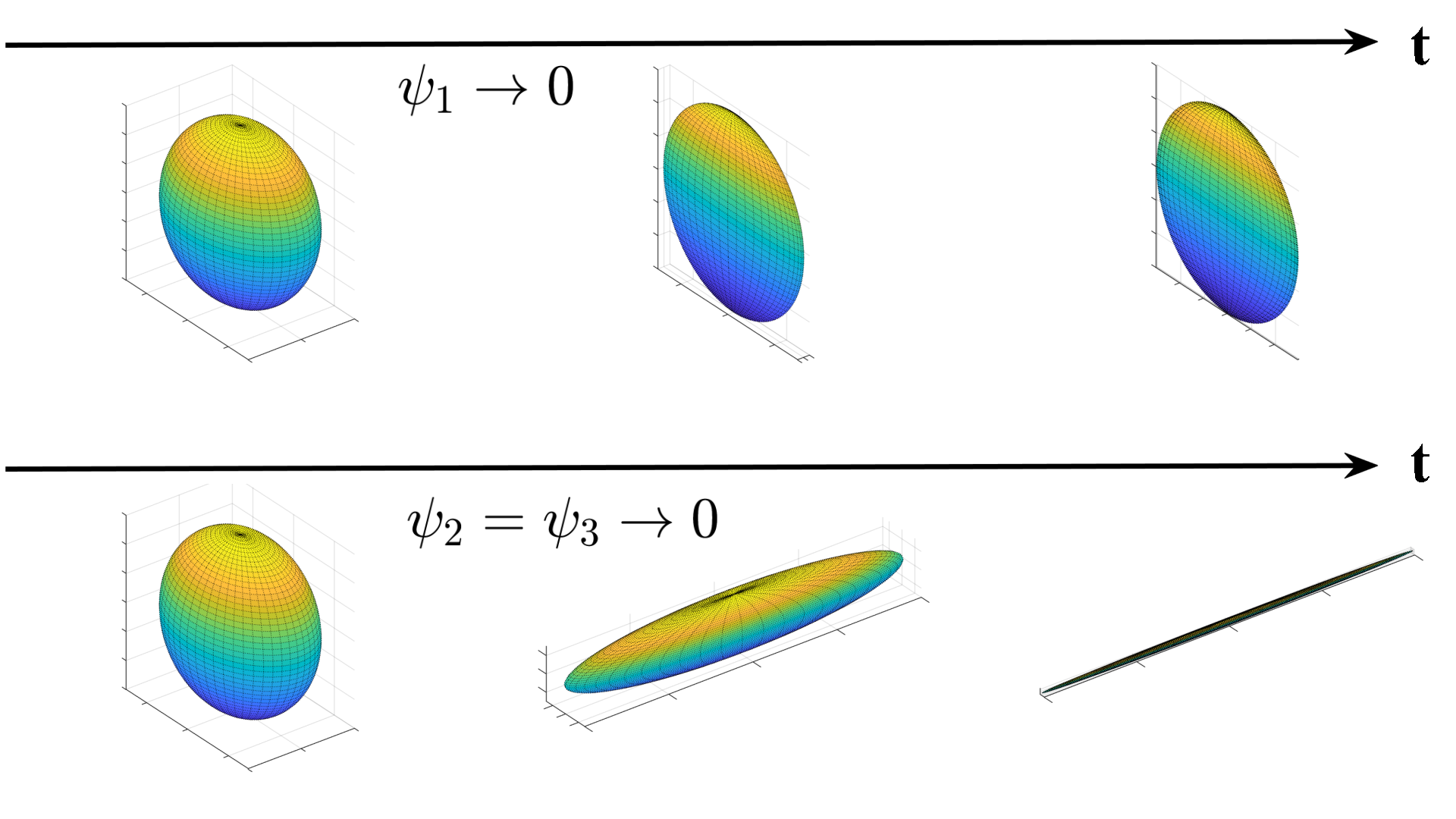}
    \caption{The ellipsoid defined by $\psi_1\psi_2\psi_3$ has a volume equal to the sphere defined by $\psi^3$ (left panel). The limits of ($\psi_1=0,\psi_2\neq 0$) and ($\psi_1\neq 0 ,\psi_2=0$) are ill defined in terms of $(\beta,\psi)$ as they no longer represent a 3-dimensional object (right panel). }
    \label{fig:sphere_and_ellips}
\end{figure}

We study the geometric meaning of transformation given in eq.~\eqref{eq:Transformation}. We first look at the geometric meaning of the fields $\psi_1,\psi_2,\psi_3$. The fields align with the three spatial axes, and their product is an ellipsoid 3-volume which is equal to the volume of a sphere with radius $\psi$ (see figure~\ref{fig:sphere_and_ellips}). Thus $\psi$ is the radius of the 2-sphere of equal volume to the ellipsoid, and as such, is a good measure of the gauge field isotropic component. The anisotropic component of the fields is given by a double Riemann projection. Since $\psi_2=\psi_3$ by virtue of the axial symmetry, we examine the relation between $\psi_1$ and $\psi_2$.

\begin{figure}[t]
    \centering
    \includegraphics[width=0.85\textwidth]{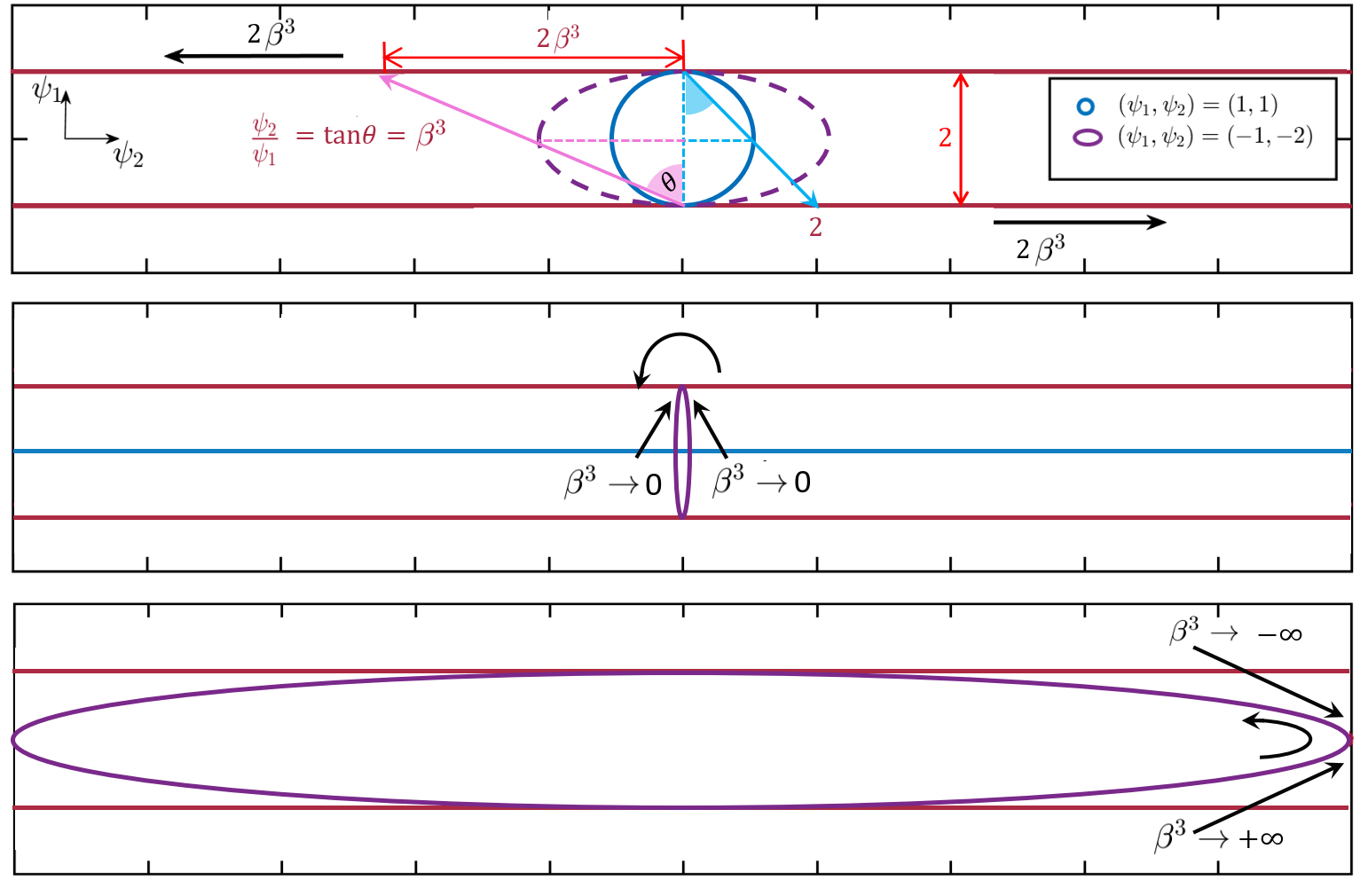}
    \caption{A double Riemann projection defined by the quantity ${\psi_2}/{\psi_1}$. In general the signs of $\psi_1$ and $\psi_2$ are not one-to-one correlated. Thus the upper and lower red lines are both possible projection lines. For clarity we normalize $|\psi_1|=1$ in this figure. The middle panel shows the case when ${\psi_2}/{\psi_1}\rightarrow 0$. In this case the evolution of $\beta(t)$ field from positive values to negative values is continuous at $\beta=0$. The bottom panel shows the case where there is an extreme zoom out of the $\psi_2$ axis. The points $\beta^3\rightarrow \infty$ and $\beta^3\rightarrow-\infty$ are geometrically identical, enabling a discontinuity and sign change of $\beta(t)$.} \label{fig:Double_Riemann}
\end{figure}

The transformation given in eq.~\eqref{eq:Transformation} implies $\beta^3 = {\psi_2}/{\psi_1}$. We suggest the following visualization: Consider the ellipse created by $\psi_1$ and $\psi_2$, such that $\psi_1$ is the vertical coordinate, and $\psi_2$ the horizontal. Now let us stretch a horizontal line `$l$' perpendicular to $(-\psi_1,0)$, and take the continuation of line connecting $\left(\psi_1,0\right)$ and $\left(0,\psi_2\right)$. The intersection of the continuation with `$l$' defines $2\beta^3$ as seen in figure~\ref{fig:Double_Riemann}.   With this visualization we conclude two things: 1) Positive values are given when $sgn(\psi_1)=sgn(\psi_2)$ and negative values when $sgn(\psi_1)=-sgn(\psi_2)$. Thus both upper and lower limiting lines are needed; and 2) Since there is no one-to-one correlation between the signs of $\psi_1$ and $\psi_2$, $\beta$ can change signs during the evolution of the system. This is contrary to the previous idea of $\beta=0$ being a hard separatrix. In practice, when using $(\psi_1,\psi_2)$ coordinates and tracking $\beta$ as derived coordinates, we see slow-roll trajectories that start at $\beta>0$ and end at $\beta=-1$.  Upon examination of $\beta$'s evolution starting from a point well within the positive-$\beta$ `no-go' region, we find two distinct behaviors: i) cases in which $\beta(t)$ continuously passes through $\beta=0$ point (see the middle panel of figure 14), and ii) cases in which $\beta(t)$ switches sign but it is discontinuous at $\beta=0$ point (the bottom panel). By this geometric representation it is now clear that the limit $\beta\rightarrow\infty$ is identified with $\beta\rightarrow-\infty$. So it makes sense to observe trajectories where $\beta\rightarrow\infty$ `jump' to extremely negative values (i.e. $\beta\rightarrow-\infty$) and go to values of negative $\beta$ of order $\mathcal{O}(1)$. Another behaviour that was tracked is the oscillations around $\beta=0$ on some limiting cycle in $(\beta,\dot{\beta})$ until eventually converging to either $\beta=1$ or $\beta=-1$.

\section{Cosmic no-hair conjecture and axion-SU(2) models}
\label{sec-cosmic-no-hair}

The Universe at cosmological scales looks homogeneous and isotropic.
Given that the
cosmic evolution may start from a generic initial condition over which we have no control, it is natural to seek a dynamical explanation, i.e., isotropic and homogeneous Universe is an
attractor of the cosmic evolution.  The first such attempt, called ``cosmic no-hair conjecture'', was made in \cite{Gibbons:1977mu,Hawking:1981fz}
arguing that the late-time behavior of any accelerating Universe is an isotropic Universe.
Before our current work, it was shown that the axion-SU(2) inflation system satisfies the cosmic no-hair conjecture, i.e., anisotropies are always diluted by inflation within a few e-folds. However, it was based on restrictive conditions: i) Bianchi type-I geometry with axial symmetry, where ii)  anisotropies of the metric and gauge fields are diagonal in the same frame. In this section, we prove the generality of this result in Bianchi type-I geometry given by
\begin{equation}
    ds^2=-dt^2 + e^{2\alpha(t)} e^{2\sigma_{ij}(t)} dx^idx^j ,
\end{equation}
where $\sigma_{ij}(t)$ is a generic symmetric and traceless matrix.

The most generic Bianchi type-I \textit{homogeneous} $SU(2)$ gauge field configuration in temporal gauge ($A^a_0=0$) can be written as
\bea
A^a_{i}(t) = \psi(t)~ e^{\beta_{ij}(t)} e^a_{j}(t) = \bar{A}(t) ~ \delta^a_j \times \bigg( e^{\beta(t)+\sigma(t)}\bigg)_{ij},
\eea
where $\bar{A}(t)$ is a shorthand for the isotropic part of the gauge field as
\bea
\bar{A}(t) =\psi(t) e^{\alpha(t)},
\eea
and $\beta_{ij}(t)$ is a symmetric traceless $3\times 3$ matrix to quantify anisotrpy of the gauge field.
Here, $\sigma_{ij}$ and $\beta_{ij}$ matrices parametrize 
the deviation from isotropy in the geometry and gauge field configuration, respectively. In the following we study these homogeneous fields with the theory
\bea\label{LA}
\mathcal{L}_A = - \frac12 {\rm{Tr}}[F_{\mu\nu} F^{\mu\nu}]-\frac{\lambda \chi}{2f} {\rm{Tr}}[F_{\mu\nu} \tilde{F}^{\mu\nu}].
\eea

The Yang-Mills term associated with the above homogeneous field configuration is
\bea
- {\rm{Tr}}[F_{\mu\nu} F^{\mu\nu}]=\p_t (\bar{A} e^{\beta+\sigma})_{ik} \p_t (\bar{A} e^{\beta+\sigma})_{kj} h^{ij}&&\\ \nonumber - \frac12 g_{ \scriptscriptstyle{A}}^2\bar{A}^4 e^{-4\alpha} &{\bigg( {\rm{Tr}}[e^{\beta+\sigma}.e^{\beta+\sigma} ]^2 - {\rm{Tr}}[e^{\beta+\sigma}.e^{\beta+\sigma}.e^{\beta+\sigma}.e^{\beta+\sigma} ] \bigg)}&.~~~~
\eea
Note that both the anisotropic parts of the geometry and gauge fields contribute to the Yang-Mills term.
 The Chern-Simons term can be written as
\bea 
&& \frac{\chi}{4} {\rm{Tr}}[F_{\mu\nu} \tilde{F}^{\mu\nu}]= \frac14 \frac{\chi }{\sqrt{-g}} \epsilon^{\mu\nu\lambda\sigma} \p_{\mu} \bigg[ A^a_{\nu} \p_{\lambda} A^a_{\sigma} + \frac{2}{3} g_{ \scriptscriptstyle{A}} \epsilon^{abc} A^a_{\nu} A^b_{\lambda} A^c_{\sigma}\bigg]\nonumber\\
&& =  g_{ \scriptscriptstyle{A}} e^{-3\alpha} \chi ~ \p_t \bigg[ \bar{A}^3 e^{{\rm{Tr}}[\beta+\sigma]} \bigg] = \frac{\chi}{4} {\rm{Tr}}[F_{\mu\nu} \tilde{F}^{\mu\nu}]_{\rm isotropic}. \label{cosmic-hair}
\eea
In the last equations we used the fact that $\beta_{ij}$ and $\sigma_{ij}$ are both traceless. Note that the anisotropies do not make any contribution to the Chern-Simons. 

The above discussion implies that the axion field only sources the isotropic part of the gauge field. However, the anisotropic parts remain sourceless as in pure Yang-Mills theory without the Chern-Simons term. Thus, we conclude that cosmological models with gauge field theory given in eq. \eqref{LA} satisfy the cosmic no-hair conjecture. 

\section{Conclusions}\label{sec:conclusions} 
  Gauge fields may survive the exponential expansion of the Universe during inflation. If such a gauge field acquires a VEV, it might pose a threat to the spatial isotropy at cosmological scales. Within the SU(2)-axion inflation models in Bianchi type-I geometry, in which the gauge field is massless, the isotropic solution is the attractor, and the system isotropizes in a few e-folds.  The stability of the isotropic gauge field solution in the spectator SU(2)-axion model was studied in \cite{Wolfson:2020fqz}. It was shown that there exist parts of the phase space, where $\dot{A}_{\mu}\gg H_0 A_{\mu}$, in which the numerical solution is not stable, the so-called no-go area.    The aim of this paper was to study this system with a different parametrization to investigate the nature of the no-go area in detail.
  
  We found that the numerical breakdown observed in \cite{Wolfson:2020fqz} was an artifact of performing the numerical analysis based on the $(\psi,\beta)$ parametrization, which is not well-defined at $\beta=0$ and unstable at $\beta\rightarrow\pm\infty$. This is problematic in the no-go area, i.e., large gauge field's kinetic term, in which the $\beta(t)$ field switches sign during its evolution. Using the $(\psi_1,\psi_2)$ parametrization, which is well-defined throughout the phase space, we revisited the previous analysis. We found that the points in the no-go area also isotropize within a few e-folds. Meanwhile, the gauge field's VEV either dilutes away, or the isotropic configuration is the attractor solution.\\
  
~The focus of the current work was to prove the stability of the isotropic background in the spectator axion-SU(2) model. Given the importance of particle production in axion-inflation models with gauge fields,  we have a quick qualitative discussion on this issue. The extensive study of this effect requires a separate study which we relegate to future work. Once the gauge field-axion system is coupled to matter fields, it produces particles during inflation via the Schwinger effect \cite{Maleknejad:2018nxz,Lozanov:2018kpk,Domcke:2018gfr,Maleknejad:2019hdr,Mirzagholi:2019jeb}. Assuming an isotropic SU(2) VEV as the source of the Schwinger effect, the induced scalar and fermionic currents are also direction independent (see e.g. eq. (4.3) in \cite{Lozanov:2018kpk} and eq. (62) in \cite{Mirzagholi:2019jeb}). Conversely, in this work the gauge field's VEV is not exactly isotropic. However, its deviation from the isotropic configuration decays exponentially fast. During the short period before the isotropization, the gauge field's VEV is larger in one direction. Therefore, we expect that in addition to the above isotropic current, this anisotropy will induce a temporary anisotropic current in its given direction. This effect is roughly similar to the case studied in \cite{Kobayashi:2014zza} (see eq. (2.44)). But in our case, once the system is isotropized, this anisotropic particle production stops.\\

Before our current work, it was analytically shown that the axion-SU(2) inflation system satisfies the cosmic no-hair conjecture, i.e., anisotropies always dilute away within a few e-folds. However, it was based on restrictive conditions: i) Bianchi type-I geometry with axial symmetry, where ii)  anisotropies of metric and gauge field are diagonal in the same frame. In this work, we proved the generality of this result.  More precisely, the axion only sources the isotropic part of the gauge field's VEV (see eq. \eqref{cosmic-hair}). Therefore, all (massless) SU(2)-axion models in Bianchi type-I geometry satisfy the cosmic no-hair conjecture.

%----------------------------------------------%
\acknowledgments
I. W. would like to thank the Max Planck Institute for Astrophysics for the generous hospitality and resources made available during this research. 
The work of E.K. was supported in part by JSPS KAKENHI Grant No.~JP20H05850 and JP20H05859, and the Deutsche Forschungsgemeinschaft (DFG, German Research Foundation) under Germany's Excellence Strategy - EXC-2094 - 390783311.
The work of T.K. was supported in part by JSPS KAKENHI Grants No.~JP20H04745 and No.~JP20K03936.

\appendix

\section{Equations for numerical integration\label{Appendix:A}}
For the sake of making these results reproducible, we write the equations of motion and the integration process in detail. 
Due to the complexity of the equations themselves however, we write the general process of deriving and solving those, rather than explicit terms.

\subsection{Equations of motion for axion and inflaton}
In the spectator case, we add an inflaton sector which is minimally coupled to the axion-SU(2) sector. Thus the equation of motion for the inflaton is the usual Klein-Gordon equation:
\begin{gather}
    \ddot{\phi}+3\dot{\alpha}\dot{\phi}+\frac{dV}{d\phi}=0.
\end{gather}
The equation of motion for the axion is more complex but is still separable from the other coordinates:
\begin{gather}
    \ddot{\chi}+3\dot{\alpha}\dot{\chi}=-\frac{g\lambda\psi_1}{f}\left[2\psi_2\dot{\psi}_1 +\psi_1\left(3\dot{\alpha}\psi_2+\dot{\psi}_2\right)\right]+\mu^4 \sin\left(\frac{\chi}{f}\right).
\end{gather}
The two equations above are integrated as-is.
\subsection{Equations of motion for the gauge field and the e-folding number}
For the other coordinates we first formulate the coordinate momentum as the following:
\begin{gather}
    \Pi_q = (\dot{\psi_1},\dot{\psi_2},\dot{\alpha})\left(\begin{array}{c}
    \Pi_{q\psi_1}\\ 
    \Pi_{q\psi_2}\\ 
    \Pi_{q\alpha}
    \end{array}\right) +C_q(q',\alpha,\chi,\dot{\chi},\phi,\dot{\phi}),\label{decomposition}
\end{gather}
where $q$ stands for one of the coordinates $\psi_1,\psi_2$.
Since the coordinate momentum is defined as
\begin{align}
    \Pi_q \equiv \frac{\partial\mathcal{L}_m}{\partial\dot{q}},
\end{align}
the terms in eq.~\eqref{decomposition} are at most linear in coordinate velocity and the terms $\Pi_{qq'}$ are functions of the coordinates $q'$ alone, making the above decomposition useful.
The equation of motion for the $q$ coordinate can be stated as:
\begin{gather}
    \frac{d}{dt}\left(\sqrt{-g}\Pi_q\right)-\frac{\partial\mathcal{L}_m}{\partial q}=0.
\end{gather}
which can be simplified and decomposed into:
\begin{gather}
    (\ddot{\psi_1},\ddot{\psi_2},\ddot{\alpha})\left(\begin{array}{c}
    \Pi_{q\psi_1}\\ 
    \Pi_{q\psi_2}\\ 
    \Pi_{q\alpha}
    \end{array}\right)+(\dot{\psi_1},\dot{\psi_2},\dot{\alpha})\left(\begin{array}{c}
    \dot{\Pi}_{q\psi_1}\\ 
    \dot{\Pi}_{q\psi_2}\\ 
    \dot{\Pi}_{q\alpha}
    \end{array}\right)+3\dot{\alpha}\Pi_q +\dot{C}_q(q',\dot{q}',\alpha,\dot{\alpha}...)-\frac{\partial\mathcal{L}_m}{\partial q}=0. \label{Pis}
\end{gather}
This can be compactly represented as:
\begin{gather}
    \ddot{V}_i \Pi_{q,i} + \dot{V}_i \dot{\Pi}_{q,i} +3\dot{\alpha}\dot{V}_i \Pi_{q,i} +\dot{C}_q -\frac{\partial\mathcal{L}}{\partial q}=0,
\end{gather}
where $\bf V$ is $(\psi_1,\psi_2,\alpha)$, and ${\bf \Pi_q} = (\Pi_{q\psi_1},\Pi_{q\psi_2},\Pi_{q\alpha})$.

For the e-folding parameter $\alpha$ we derive the Friedman equation (eq. \eqref{Fried1}) with respect to time.
The Friedmann equation is
\begin{gather}
    3\dot{\alpha}^2-3\dot{\sigma}^2 = \rho_{A} +\rho_\chi +\rho(\phi),
\end{gather}
where 
\begin{align}
    \dot{\sigma}=\frac{\dot{\alpha}\left(\psi_1^2-\psi_2^2\right) +\dot{\psi_1}\psi_1-\dot{\psi_2}\psi_2}{3+2\psi_1^2 +\psi_2^2},
\end{align}
\begin{align}
    \rho_{A}= \psi_2^2\left(\dot{\alpha}+\dot{\sigma}+\tfrac{\dot{\psi_2}}{\psi_2}\right)^2 + \frac{\psi_1^2}{2}\left(\dot{\alpha}-2\dot{\sigma}+\tfrac{\dot{\psi_1}}{\psi_1}\right)^2 +\frac{g_{ \scriptscriptstyle{A}}\psi_2^2}{2}\left(2\psi_1^2 + \psi_2^2\right),
    \end{align}
    and
    \begin{align}
    \rho_{\chi}=\frac{\dot{\chi}^2}{2}+\mu^4\left(1+\cos{\tfrac{\chi}{f}}\right)\;\;,\; \rho(\phi)=\frac{\dot{\phi}^2}{2}+V(\phi).
    \end{align}
So we can reformulate this equation as
\begin{align}
    \dot{\alpha}^2 = \dot{V}_i \dot{V}_j A_{ij} +C_{\alpha}(\alpha,q,...),
\end{align}
where $\hat{A}$ is a $3\times3$ matrix that does not include $\dot{\psi}_1,\dot{\psi}_2,\dot{\alpha}$.
After deriving w.r.t time we have:
\begin{gather}
    2\dot{\alpha}\ddot{\alpha}= 2\ddot{V}_i \dot{V}_j A_{ij}+ \dot{V}_i\dot{V}_j \dot{A}_{ij} +\dot{C}_{\alpha}.
\end{gather}
We can redefine $\hat{A}$ to absorb the term on the left as $\hat{\tilde{A}}$ to yield the following from the $\alpha$ acceleration equation
\begin{gather}
    \ddot{V}_i\dot{V}_j \tilde{A}_{ij} =\frac{ -\dot{V}_i \dot{V}_{j} \dot{A}_{ij} -\dot{C}_{\alpha}}{2}.
\end{gather}
\hfill \break
~Finally, we construct the $3\times 3$ matrix 
\begin{gather}
    M_{ij}= \left\{\begin{array}{lr}
        \dot{V}_k \dot{\tilde{A}}_{ik}& q_j=\alpha \\
        \Pi_{q_ji}& q_j=\psi_1,\psi_2
        \end{array}\right. ,
\end{gather}
and the vector
\begin{align}
    U_j = \left\{\begin{array}{lcr}
         -\frac{1}{2}\left(\dot{V}_i \dot{V}_{k} \dot{A}_{ik} +\dot{C}_{\alpha}\right)       &\;\hspace{1em}& q_j=\alpha\\
         &&\\
       -\dot{V}_i \dot{\Pi}_{q_j i}-3\dot{\alpha}\dot{V}_i \Pi_{q_j i} -\dot{C}_{q_j}+\frac{d\mathcal{L}}{dq_j}         &\;\;\;\;\;\;\;\;\;\;\;& q_j=\psi_1,\psi_2
    \end{array}\right. .
\end{align}
Thus the equation we have is now simply constructed as
\begin{align}
    \ddot{V}_i M_{ji} = U_j, 
\end{align}
to which we have a solution 
\begin{align}
    \ddot{V_i}= U_j M^{-1}_{ji},
\end{align}
as long as $M$ is regular. The same scheme can be employed in the $(\beta,\psi)$ coordinate system. However, in that coordinate system $M$ sometimes becomes either non-regular or otherwise poorly scaled such that numerical errors become a critical issue. This accounts (numerically) for the previous notion of a no-go region.

Another benefit of this method is that $\dot{\alpha}$ can be derived in two mathematically equivalent, but numerically complementary ways. We thus take the result of the above integration for $\int\ddot{\alpha}dt=\dot{\alpha}$, and compare it with the result of the explicit Friedmann equation $3\dot{\alpha}^2=3\dot{\sigma}^2 +\rho$. This allows us to monitor the numerical error and better restrict it.

\end{document}